\documentclass[a4paper]{article}
\usepackage[centertags]{amsmath}
\usepackage{amssymb}
\newcommand{\qq}{\begin{equation}}
\newcommand{\qqq}{\end{equation}}
\newcommand{\dd}{{\mathrm{d}}}
\newcommand{\RR}{{\mathbb{R}}}

\newcommand{\ZZ}{{\mathbb{Z}}}

\newcommand{\EE}{{\mathbb{E}}}
\newcommand{\pp}{{\hspace{0.03cm}\prime}}

\newcommand{\secref}[1]{{Sect.~\ref{#1}}}

\newcommand{\appref}[1]{{Appendix~\ref{#1}}}
\newcommand{\tabref}[1]{{Table~\ref{#1}}}

\newcommand{\dimface}{\mathsf}

\begin{document}

\title{\Large\bf Sticky behavior of fluid particles
in the compressible Kraichnan model}
\author{\ \\ Krzysztof Gaw\c{e}dzki$^{*,1}$\ \ and\ \ 
P\'{e}ter Horvai$^{1,2}$
\\ }

\maketitle

\ 
\vskip -1.4cm
\

\begin{center}
$^{*}$\,member of C.N.R.S. \\
$^{1}$Laboratoire de Physique, ENS-Lyon, 46 All\'{e}e d'Italie,
69364 Lyon, France\\
$^{2}$Centre de Physique Th\'{e}orique,
\'{E}cole Polytechnique, 91128 Palaiseau, France
\end{center}
\vskip 1cm

\begin{abstract}
We consider the compressible Kraichnan model of turbulent advection
with small molecular diffusivity and velocity field regularized at
short scales to mimic the effects of viscosity.  As noted in
\cite{EVDE1}, removing those two regularizations in two opposite
orders for intermediate values of compressibility gives Lagrangian
flows with quite different properties.  Removing the viscous
regularization before diffusivity leads to the explosive separation of
trajectories of fluid particles whereas turning the regularizations
off in the opposite order results in coalescence of Lagrangian
trajectories.  In the present paper we re-examine the situation first
addressed in \cite{EVDE2} in which the Prandtl number is varied when
the regularizations are removed. We show that an appropriate
fine-tuning leads to a sticky behavior of trajectories which hit each
other on and off spending a positive amount of time together.  We
examine the effect of such a trajectory behavior on the passive
transport showing that it induces anomalous scaling of the stationary
2-point structure function of an advected tracer and influences the
rate of condensation of tracer energy in the zero wavenumber mode.
\end{abstract}
\vskip -0.3cm
\

\tableofcontents

\ 
\vskip -0.58cm
\

\section{Introduction}
\label{sect:intro}

Certain qualitative features of transport of scalar or vector
quantities by highly turbulent flows may be understood using
drastically simplified models where the back-reaction of the
transported quantity on the velocity dynamics is ignored (passive
advection) and the hydrodynamical velocities are replaced by a simple
synthetic random velocity ensemble.  In recent years much attention
has been attracted by the study of passive advection in a Gaussian
ensemble of time-decorrelated velocities, first considered by
Kraichnan \cite{Kr68} and Kazantzev \cite{Ka68}.  The model has led to
a new understanding of the statistical intermittency of advected
quantities and of the interplay between fluid compressibility and
transport properties, see \cite{FGV} and references therein.  The
passive advection by a $d$-dimensional velocity field
$\vec{v}(t,\vec{r})$ is intimately related to the behavior of the
Lagrangian trajectories (i.e.\,\,trajectories of fluid particles)
perturbed by noise.  Such trajectories satisfy the stochastic ordinary
differential equation
\begin{equation}
\label{eq:adv-stoch-diffus}
  \frac{\dd}{\dd t} \vec{R}(t)
=
  \vec{v}(t,\vec{R}(t))
  + \sqrt{2\kappa} \, \vec{\eta}(t),
\end{equation}
where $\vec{\eta}(t)$ is a $d$-dimensional white noise independent of
velocity (and of noises of other particles) and $\kappa>0$ is the
(molecular) diffusivity.  The advection of a passive tracer
$\theta(t,\vec{r})$ is governed by the advection-diffusion equation
\begin{equation}
\label{eq:ad-dif}
  \partial_t \theta +
  \vec{v}\cdot\vec{\nabla} \theta -
  \kappa{\vec{\nabla}}^2 \theta
=
  f\,,
\end{equation}
where $f(t,\vec{r})$ is a source or forcing term.  In the absence of
forcing (i.e.\,\,for $f=0$) the passive tracer is carried along the
particle trajectories so that
\begin{equation}
\label{eq:scevol}
  \theta(t,\vec{r})
=
  \overline{\theta(0,\vec{R}(0;t,\vec{r}))}
\,.
\end{equation}
Here $\vec{R}(t';t,\vec{r})$ denotes a particle trajectory that passes
through point $\vec{r}$ at time $t$ and the overline stands for the
average with respect to the noise $\vec{\eta}(t)$.  Note that the
forward tracer evolution is described by Lagrangian trajectories going
backwards in time.  Similarly, in the presence of the source, the
tracer evolves according to the equation
\begin{equation}
\label{eq:scevolf}
  \theta(t,\vec{r})
=
  \overline{\theta(0,\vec{R}(0;t,\vec{r}))} +
  \int_0^t \overline{f(s,\vec{R}(s;t,\vec{r}))} \,\dd s
\,.
\end{equation}
This way, in random velocities, the statistics of advected quantities
like $\theta$ is linked to the statistical properties of (noisy)
trajectories.  The present article exploits this relation 
in yet another situation in the Kraichnan model of turbulent advection.

\subsection{The Kraichnan model}

The Kraichnan model describes advection by a stochastic velocity field
$\vec{v}(t,\vec{r})$ with Gaussian mean-zero statistics, stationary
and decorrelated in time, homogeneous and isotropic in space and with
fixed compressibility degree $\wp$ characterizing the relative
strength of the incompressible and potential components of the
velocity.  As in any centered Gaussian ensemble, the statistics of
velocities is totally characterized by the 2-point function
$\langle\vec{v}(t,\vec{r})\otimes\vec{v} (t',\vec{r}^\pp)\rangle$.  In
what follows, we shall need the statistics of the equal-time velocity
differences that is determined by the reduced expression
\begin{equation}
\label{def:structure-v}
\begin{split}
 \frac{_1}{^2}\, \langle
    (\vec{v}(t,\vec{r}) - \vec{v}(t',\vec{r}^\pp))^{\otimes 2}
  \rangle 
&=
  \delta(t-t') \,
  D_0 \int \frac{1 - e^{i \vec{k} . (\vec{r}-\vec{r}^\pp)}}
                {|\vec{k}|^{d + \xi}}
           \vec{\vec{P}}(\vec{k},\wp)
           \frac{\dd \vec{k}}
                {(2\pi)^d}
\\
&\equiv  
  \delta(t-t')\,\vec{\vec{d}}(\vec{r}-\vec{r}^\pp).
\end{split}
\end{equation}
The rank 2 tensor $\vec{\vec{d}}(\vec{r}^\pp-\vec{r})$ is the second
order spatial structure function of the velocity field.  The rank 2
tensor $\vec{\vec{P}}(\vec{k},\wp)$, invariant under rotations and of
trace 1, is taken to be
\begin{equation}
  P_{ij}(\vec{k},\wp)
= \frac{1-\wp}{d-1} \left( \delta_{ij} - \frac{k_i k_j}{k^2} \right)
  + \wp \frac{k_i k_j}{k^2}
\nonumber
\end{equation}
with $0 \leq \wp \leq 1$. The coefficient $\wp$ is equal to
$
 \frac{\partial_i \partial_j d_{ij}(\vec{r})}
       {\partial_i \partial_i d_{jj}(\vec{r})}
$
(summation convention!) for all $\vec{r}\not=0$, so that it is
meaningful to characterize $\wp$ as the ratio $\,\frac{\langle
(\partial_i v_i)^2 \rangle}{\langle (\partial_i v_j)^2 \rangle}\,$
even though the velocity field is not smooth for $\xi < 2$.  The value
$\wp = 0$ corresponds to an incompressible velocity field, whereas
$\wp = 1$ to a potential (e.g.\ irrotational) one.  Note however that
if the physical space is one dimensional (case $d=1$) then necessarily
$\wp = 1$ because a one-dimensional velocity field is always
potential. One simply takes $P=1$ in this case.  The spatial velocity
structure function $\vec{\vec{d}}(\vec{r})$ scales in $|\vec{r}|$ with
power $\xi$.  We shall take $\xi$ between 0 and 2.  This choice
guarantees that the typical velocity realizations are non-Lipschitz.
More exactly, they are H\"{o}lder-continuous with exponent
$\frac{1}{2}\xi'$ for any $\xi'<\xi$ (the realistic turbulent
velocities are believed to have H\"{o}lder exponent $\leq\frac{1}{3}$
in the limit of infinite Reynolds numbers \cite{Onsager}).  Note that
the (positive) constant $D_0$ in \eqref{def:structure-v} has the
dimension $\frac{{(\dimface{length})}^{2-\xi}}{\dimface{time}}$.

\subsection{Regularizing effects}

Real flows are always regularized at small scales by viscous effects.
That is, the velocity field is smooth and the power law scaling with
$\xi < 2$ is observed only at distances much larger than the {\bf
viscous cutoff scale} $l_\nu$ that becomes very small only for high
Reynolds numbers (if the integral scale of turbulence is kept fixed).
The small $\vec{r}$ behavior $\propto|\vec{r}|^\xi$ of
$\vec{\vec{d}}(\vec{r})$ above comes from the the slow decay at large
$\vec{k}$ of its Fourier transform (the term under the integral sign
in \eqref{def:structure-v}).  We shall mimic the effect of viscosity
in the present model by introducing an ultraviolet regulator and
replacing $\vec{\vec{d}}(\vec{r})$ by
\begin{equation}
\label{def:d-reg}
  \vec{\vec{d}}(\vec{r};l_\nu)
\equiv
  D_0 \int \frac{1 - e^{i \vec{k} . \vec{r}}}
                {|\vec{k}|^{d + \xi}}
           \vec{\vec{P}}(\vec{k},\wp)
           f(l_\nu |\vec{k}|)
           \frac{\dd \vec{k}}
                {(2\pi)^d}
\end{equation}
where the function $f$ is taken positive, smooth, decreasing, fast
decaying at infinity and with $f(0) = 1$, $f'(0) = 0$. The viscosity
$\nu$ itself may be defined as
\begin{equation}
\label{eq:visc}
\nu=D_0 l_\nu^\xi
\end{equation}
which is the only combination of $D_0$ and $l_\nu$ of dimension
$\frac{(\dimface{length})^2}{\dimface{time}}$.
\vskip 0.15cm

Another regularizing effect in real flows comes from finite molecular
diffusivity.  The tracer field $\theta(t,\vec{r})$ passively
transported by the flow will be smoothed by diffusion, see
\eqref{eq:scevol}. In typical velocities, this smoothing effect
becomes important at scales much smaller than the {\bf diffusive
cutoff scale} $l_\kappa$.  Beneath this scale diffusion dominates
advection.  We modeled molecular diffusivity by adding white noise
terms to the fluid particle velocities (independent ones for different
particles, see \eqref{eq:adv-stoch-diffus}).  In the Kraichnan model,
the scale $l_\kappa$ beneath which diffusion dominates advection may
be expressed in terms of $\kappa$, $D_0$ and $l_\nu$, with
$l_\kappa=0$ corresponding to $\kappa=0$.  The expression may be
inverted to calculate $\kappa$ in terms of $D_0,l_\kappa,l_\nu$.
Later on we shall specify such a relation in the case when $l_\kappa
\ll l_\nu$, see \eqref{expr:kappa}.  This will provide an expression
for the Prandtl number $\mathit{Pr} \equiv \frac{\nu}{\kappa}$.

\subsection{Statistics of fluid particles}

Even in a simple random velocity ensemble, the statistics of the
Lagrangian flow may be quite complicated.  It may be studied by
looking at the joint $N$-particle probability density functions (PDFs)
defined by
\begin{equation}
\label{eq:Npart}
  P_N^{t',t} (\vec{r}_{1},\dots,\vec{r}_{N};
              \vec{r}^\pp_1    ,\dots,\vec{r}^\pp_N     )
\ =\ 
  \Big\langle \prod\limits_{n=1}^N
    \overline{\delta(\vec{r}^\pp_n-\vec{R}(t;t',\vec{r}_{n}))}
  \,\Big\rangle
\,.  
\end{equation}
Here, as before, the overline denotes the average over the
(independent) white noises and $\langle\,\cdots\rangle$ stands for the
velocity ensemble average.  In the Kraichnan model, due to the
temporal decorrelation of velocities, the PDFs \eqref{eq:Npart} are
Markovian and they define a consistent hierarchy of $N$-particle
stationary Markov processes that contains the complete information
about the statistics of the Lagrangian flow and of the velocities
themselves \cite{LeJR2}.  In this paper we shall be interested
uniquely in the behavior of the separation of a pair of Lagrangian
particles.  The main object of our interest will be the PDF of finding
their time $t$ separation equal to $\vec{r}$, given that their time
$t'$ separation is equal to $\vec{r}_0$:
$$
  P^{t,t'} (\vec{r},\vec{r}^\pp)
=
  \left\langle
    \overline{\delta(\vec{r}^\pp - R(t';t,\vec{r}) + R(t';t,\vec{0}))}
  \right\rangle.
$$
In stationary, time-reversal invariant velocity ensembles, such as the
Kraichnan model, $P^{t,t'}$ depends only on $|t-t'|$ and we shall use
the notation $P^{|t-t'|}$.
\vskip 0.15cm

The PDF $P^{t,t'}(\vec{r},\vec{r}^\pp)$ governs the free decay of the
2-point function of the passive tracer, evolving according to
equation \eqref{eq:ad-dif}%
\footnote{
        In the Kraichnan velocity that is white in time,
        Eq.~\eqref{eq:ad-dif} should be interpreted as a Stratonovich
        stochastic differential equation, see e.g.\ Sect.\ II.C.3 of
        \cite{FGV}.
}
with $f=0$.  If at time zero the tracer is distributed independently
of the velocity field and the trajectory noises, with a homogeneous
2-point function $\langle \theta(0,\vec{r}) \, \theta(0,\vec{r}^\pp)
\rangle = F(0, \vec{r}^\pp-\vec{r})$, then at a later time $t$ its
2-point function is given by
\begin{equation}
\label{eq:2ptsc}
  F(t,\vec{r})
=
  \int
    F(0,\vec{r}^\pp)\,P^{t,0}(\vec{r},\vec{r}^\pp)
  \,\dd\vec{r}^\pp
\end{equation}
as follows from \eqref{eq:scevol}.
Similarly, if the scalar source $f$ is a random field with mean zero
and 2-point function
\begin{equation}
\label{eq:fc}
  \langle f(t,\vec{r})\,f(t',\vec{r}^\pp) \rangle
=
  \delta(t-t') \, \chi(\vec{r}-\vec{r}^\pp)
\end{equation}
and it is independent of the velocity field, the trajectory noises and
the initial scalar distribution, then the evolution of the tracer
2-point function is described by the relation
\begin{equation}
\label{eq:2ptscf}
  F(t,\vec{r})
=
  \int F(0,\vec{r}^\pp)\, P^{t,0}(\vec{r},\vec{r}^\pp) \,\dd\vec{r}^\pp
  +\int_0^t \dd s
     \int\chi(\vec{r}^\pp)\,P^{t,s}(\vec{r},\vec{r}^\pp) \,\dd\vec{r}^\pp
\end{equation}
as follows from \eqref{eq:scevolf} by taking averages.
\vskip 0.15cm

Our aim is to describe the particle pair separation in Kraichnan
velocities at large scales, much larger than the cutoff scales $l_\nu$
and $l_\kappa$.  The effective description will depend on the Prandtl
number and on the scales involved.  Instead of maintaining the cutoffs
finite, we want to give an effective large scale theory without
cutoffs but with some specific boundary condition at vanishing
separation.  Such a theory should give rise to the same behavior at
the large scales of interest.  An alternative way of doing this is by
taking the cutoffs $l_\nu$ and $l_\kappa$ to zero in a fashion that
preserves large scale behavior and to examine the boundary condition
that arises in this limiting process.

\subsection{Lagrangian dispersion}

As first derived in \cite{Kr68}, the time evolution of the scalar
2-point function of the tracer in the Kraichnan ensemble of velocities
is governed by the differential equation
\begin{equation*}
  \partial_t F
=
  \mathcal{M}_{\nu,\kappa} F
\end{equation*}
where $\mathcal{M}_{\nu,\kappa}$ is a partial differential operator
which can be written in terms of the velocity structure function
$\vec{\vec{d}}\,$ and diffusivity $\kappa$ as
\begin{equation*}
  \mathcal{M}_{\nu,\kappa}
\equiv
  d_{ij}(\vec{r};l_\nu) \partial_i \partial_j +
  2 \kappa {\vec{\nabla}}^2.
\end{equation*}
It follows that the separation PDF takes the heat kernel form
\begin{equation}
\label{eq:heatk}
  P^{t}(\vec{r}_0,\vec{r})
=
  {\rm e}^{\mathcal{M}_{\nu,\kappa}} (\vec{r}_0,\vec{r}).
\end{equation}
As we shall be only interested in distances between two particles,
usually called the {\bf Lagrangian dispersion}, and not in the angular
distribution of the particle separation, we may project
\eqref{eq:heatk} to the rotationally invariant sector.  This is done
by restricting the action of operator $\mathcal{M}_{\nu,\kappa}$ to
functions of the radial variable $r = |\vec{r}|$ only.
$\mathcal{M}_{\nu,\kappa}$ is a rotationally invariant operator and it
maps functions of $r$ to functions of $r$.  We shall denote by
$M_{\nu,\kappa}$ its restriction to such functions.  It is easy to
show that
\begin{equation}
\label{eq:riop}
  M_{\nu,\kappa} 
=
  \tilde{D}_0
  \left[ f_1(r) \partial_r^2 + f_2(r) \partial_r \right]
\end{equation}
with the coefficient functions $f_{1,2}(r)\equiv
f_{1,2}(r;l_\nu,l_\kappa)$ expressed in terms of the spatial velocity
structure function $\,\vec{\vec{d}}\,$ by
\begin{gather}
\label{eq:f_1}
\tilde{D}_0 f_1(r)
=
  d_{ij}(\vec{r};l_\nu) \frac{r_i r_j}{r^2} + 2\kappa
\\
\label{eq:f_2}
\tilde{D}_0 f_2(r)
=
  \frac{1}{r}
    d_{ij}(\vec{r};l_\nu) (\delta_{ij} - \frac{r_i r_j}{r^2})
  +
  2\frac{d-1}{r} \kappa
\end{gather}
for any $\vec{r}$ such that $|\vec{r}| = r$.  For detailed calculation
of the values of $f_1$ and $f_2$ see \appref{apx:f1-f2}.  It turns out
that, for $r$ much larger than the cut-off scales $l_\kappa,l_\nu$,
the function $f_1$ is asymptotically proportional to $r^\xi$ and the
function $f_2$ to $r^{\xi-1}$.  We choose the constant $\tilde{D}_0$
proportional to $D_0$, see \eqref{def:structure-v}, so that
$r^{-\xi}f_1(r)\to1$ when $r\to\infty$.
\vskip 0.15cm

In order to turn $M_{\nu,\kappa}$ into a true operator on the
half-line $\RR_+$, we also have to specify its domain
$\mathcal{D}(M_{\nu,\kappa})$.  This boils down to the choice of the
boundary condition to impose at zero.  The domain of
$\mathcal{M}_{\nu,\kappa}$, as long as $\kappa$ is positive, is some
locally $H^2$ Sobolev space.  For its rotationally invariant elements
this implies that their gradient vanishes at $\vec{r}=0$ resulting in
the Neumann boundary condition $\,\partial_r |_{r=0} = 0\,$ for
$M_{\nu,\kappa}$.  This way equation \eqref{eq:heatk} reduces in the
rotationally invariant sector to the heat kernel expression for the
particle distance PDF
\begin{equation}
\label{eq:half-line}
  P^t(r_0,r)
=
  {\rm e}^{t{M}_{\nu,\kappa}}({r}_0,{r}),
\end{equation}
where ${M}_{\nu,\kappa}$ is the second order differential operator on
the half-line of $r \geq 0$ given by \eqref{eq:riop}, taken with the
Neumann boundary condition at $r=0$.  We may view the right hand side
of \eqref{eq:half-line} as the transition probability density of a
Markov process $r(t)$ that describes the distance between two noisy
Lagrangian particles in the random flow and treat the underlying
diffusion process with the adapted tools, see \appref{apx:1d-dif}.
Alternatively, we may interpret \eqref{eq:half-line} as describing
kernels of a semigroup of operators and apply to it the usual analysis
toolkit.  Both methods will be developed.

\subsection{Initial classification of boundary behaviors}

The possible boundary behaviors at the origin of the
one-dimensional diffusion process $r(t)$ describing the
inter-particle distance in the absence of regularizations (i.e.\,\,for
$l_\nu,l_\kappa = 0$) may be easily classified \cite{EVDE2}.  In this
case the generator \eqref{eq:riop} of the process reduces to the
operator
\begin{equation}
\label{eq:M}
  M\equiv M_{0,0}
=
  {\tilde D}_0 r^\xi
    (\partial_r^2 + \frac{a_{\xi,\wp}}{r} \partial_r)
\end{equation}
with
\begin{equation}
\label{def:a_xi}
  a_{\xi,\wp}
=
  \frac{d+\xi}
       {1+\xi\wp}
  - 1\,,
\end{equation}
see \appref{apx:f1-f2}.  
Let us note in passing for later use that we have the implication
\begin{equation*}
  a_{\xi,\wp} < 1
\quad\Longrightarrow\quad
  a_{2,\wp} < 1
\,.
\end{equation*}
For $d \geq 2$ this is a simple consequence of \eqref{def:a_xi}.  In
the case $d=1$ recall that $\wp=1$ and then $a_{\xi,1} = 0$ for any
value of $\,\xi$.
Below, whenever the value of the compressibility degree $\wp$ is
understood, we shall write $a_\xi$ instead of the more unwieldy
$a_{\xi,\wp}$ and sometimes simply $a$ if no confusion may arise.
\vskip 0.15cm

In the case $\xi < 2$ corresponding to spatially rough flows, we can
introduce the new coordinate
\begin{equation}
\label{def:Bessel-coord}
  u
=
  u(r)
\equiv
  \frac{_2}{^{2-\xi}} {\tilde{D}_0}^{-1/2}
  r^{1 - \frac{\xi}{2}}
\,.
\end{equation}
In this coordinate the generator \eqref{eq:M} becomes
\begin{equation}
\label{expr:gen-Bessel}
  \partial_u^2 +
  \left( \frac{2a_\xi-\xi}{2-\xi} \right)
  \frac{1}{u} \partial_u
\end{equation}
and it describes the Bessel process (see \cite{HBM} for a brief {\it
r\'{e}sum\'{e}}) of {\bf parameter} $-b_\xi$ or {\bf effective
dimension} $d_{\mathrm{eff}}=2(1-b_\xi)$ where
\begin{equation}
\label{def:b_xi}
  b_\xi
=
  \frac{1-a_\xi}{2-\xi}\,.
\end{equation}
If $d_{\mathrm{eff}}$ is a positive integer then the corresponding
Bessel process describes the behavior of the norm $|\vec{W}(t)|$ of
the $d_{\mathrm{eff}}$-dimensional Brownian motion $\vec{W}(t)$.
\vskip 0.15cm

The general theory of the boundary behaviors of one-dimensional
diffusion processes was laid down by Feller in \cite{Feller}.  For the
Bessel process, the boundary behavior at zero is well known
\cite{HBM,RW} and it depends on the parameter or equivalently the
effective dimension of the process.  For $d_{\mathrm{eff}} \leq 0$,
zero is an exit boundary point (the realizations may arrive at zero in
a finite time, but they cannot come back from zero nor start there).
For $0 < d_{\mathrm{eff}} < 2$, zero is a regular boundary point (the
realizations can go to and leave zero in finite time).  For
$d_{\mathrm{eff}} \geq 2$, zero is an entrance boundary (the
realizations may start from zero, but no realization goes to zero in
finite time).  The different boundary behaviors enumerated above
correspond respectively to the regimes in the Kraichnan model that
were coined
{\bf strongly compressible} ($\wp\geq\frac{d}{\xi^2}$),
{\bf of intermediate compressibility}
($\frac{d-2}{2\xi}+\frac{1}{2}<\wp< \frac{d}{\xi^2}$)
and {\bf weakly compressible} ($\wp\leq\frac{d-2}{2\xi}+\frac{1}{2}$),
see \cite{GV,EVDE1}.
\vskip 0.15cm

In the limiting case $\xi=2$ corresponding to spatially smooth flows, we 
may introduce the new coordinate $u=\tilde{D}_0^{-1/2} \ln{r}$.  In this
coordinate the generator \eqref{eq:M} becomes
$$
  \partial_u^2+(a_2-1)\partial_u
$$
and it describes the one-dimensional Brownian motion viewed in the
frame moving with speed $-(a_2-1)$ (i.e.\,\,the Brownian motion with
a drift).  In this case $r=0$ (corresponding to $u=-\infty$) is a
natural boundary point with no realizations that attain it or leave it
in finite time.  The quantity $(a_2-1){\tilde D}_0^{1/2} =
\frac{d-4\wp}{1+2\wp} {\tilde D}_0^{1/2}$ is the Lyapunov exponent of
the Lagrangian flow and it is non-negative in the weakly compressible
regime $\wp \leq \frac{d}{4}$, and non-positive in the strongly
compressible one $\wp \geq \frac{d}{4}$, vanishing at their common
point $\wp = \frac{d}{4}$.  We sum up this classification in
\tabref{tab:b-class}.
\begin{table}
\begin{center}
\vskip 0.2cm
\begin{tabular}{|c|c|c|c|c|}
\hline
  flow & compressibility & characterization & implications & boundary \\
\hline \hline
  & weak &
  \rule[-1.4ex]{0mm}{4.1ex} $\wp\leq\frac{d-2}{2\xi}+\frac{1}{2}$ &
  $a_{\xi,\wp} \geq 1$ &
  entrance \\
\cline{2-5}
  & & & \rule[-1.4ex]{0mm}{4.1ex} $1 > a_{\xi,\wp} > \xi - 1$ & \\
  \raisebox{2ex}[0cm][0cm]{rough} &
  \raisebox{2ex}[0cm][0cm]{intermediate} &
  \raisebox{2ex}[0cm][0cm]{
    \rule[-1.4ex]{0mm}{4.1ex}
    $\frac{d-2}{2\xi}+\frac{1}{2} < \wp<\frac{d}{\xi^2}$
  } &
  \rule[-1.4ex]{0mm}{4.1ex} $1 > a_{2,\wp}$ &
  \raisebox{2ex}[0cm][0cm]{regular} \\
\cline{2-5}
  & strong &
  \rule[-1.4ex]{0mm}{4.1ex} $ \wp\geq\frac{d}{\xi^2}$ & 
  $\xi - 1 \geq a_{\xi,\wp}$ &
  exit \\
\hline
  & weak &
  \rule[-1.4ex]{0mm}{4.1ex} $\wp \leq\frac{d}{4}$ &
  $a_{2,\wp} \geq 1$ &
  \\
\cline{2-4}
  \raisebox{2.1ex}[0cm][0cm]{smooth} & strong &
  \rule[-1.4ex]{0mm}{4.1ex} $\wp\geq\frac{d}{4}$ &
  $a_{2,\wp} \leq 1$ &
  \raisebox{2.1ex}[0cm][0cm]{natural} \\
\hline
\end{tabular}
\caption{Regimes of the unregularized Kraichnan model}
\label{tab:b-class}
\end{center}
\vspace{-0.6cm}
\end{table}
The first three lines correspond to rough velocity fields with
$\xi<2$.  The last two correspond to smooth ones with $\xi=2$.  Note
that for $d \geq 4$ the weakly compressible regime extends to the
whole interval $0 \leq \wp\leq1$.
\vskip 0.15cm

The relationship between the boundary behavior of the one-dimensional
diffusion process $r(t)$ describing the distance between Lagrangian
trajectories, on the one hand, and the different regimes of the
Lagrangian flow, on the other hand, may be explained intuitively.  If
the realizations can go to zero but cannot come back, then they are
trapped there.  This implies coalescence of Lagrangian trajectories
that characterizes the strong compressibility regime, as first
observed in \cite{GV}.  If realizations can go to and come back from
zero, then one has to specify further the boundary behavior.  There
may be (and, as we find, there are indeed) different possible
behaviors of Lagrangian trajectories when they meet.  This is what
happens for intermediate compressibility, as first noted in
\cite{EVDE1}.  If realizations starting from inside the half-line can
never reach zero then no trapping is possible.  In this situation,
corresponding to weak compressibility, the realizations may, however,
enter from zero if $\xi<2$ meaning that the Lagrangian trajectories
separate explosively leading to a non-deterministic Lagrangian flow
\cite{SM}.  For $\xi=2$ the realizations can neither collapse to zero
nor explode from it.  In smooth flows, different trajectories remain
separated indefinitely in the future and the past, although they may
behave chaotically, with small separations growing exponentially in
time.

\subsection{Main result}

In any case, the regulated generator $M_{\nu,\kappa}$ of
Eq.~\eqref{eq:riop} taken with Neumann boundary condition at zero
gives rise to a diffusion process $r(t)$ on the half-line describing
the statistics of the Lagrangian dispersion.  We would like to know
how this process behaves in the limit when $l_\nu,l_\kappa \rightarrow
0$.
\vskip 0.15cm

The cases of the weakly and strongly compressible regimes where the
limiting dispersion process does not depend on the way in which the
limit is taken seem to have been treated adequately \cite{GV,EVDE1}.
In these cases only one process is possible in the limit and it
corresponds either to instantaneous reflection of trajectories when
they meet (for weak compressibility) or to coalescence of trajectories
upon the first hit (for strong compressibility).
\vskip 0.15cm

In ref.\ \cite{EVDE1,EVDE2} it was argued that, in the intermediate
regime, the possible limiting processes are described by the
unregulated generators $M$ of Eq.~\eqref{eq:M} with boundary
conditions at zero depending on the way the limit $l_\nu,l_\kappa
\rightarrow 0$ is taken.  Our analysis confirms this picture but
differs from \cite{EVDE2} in specific conclusions (the nature of the
limiting boundary conditions and the form of the limiting procedures
that lead to them).  Upon the fine-tuning \eqref{def:lambda} of
$l_\nu$ and $l_\kappa$, we find in the limit the so called
{\bf sticky} or
{\bf slowly reflecting} boundary conditions \cite{HBM,RY} parametrized
by the amount of ``glue'' $\lambda \in [0,\infty]$.
\vskip 0.15cm

The sticky behavior can be pictured intuitively in the following way.
For simplicity let us consider just the case of Brownian motion on the
half-line.  Concerning ordinary reflected Brownian motion, it is a
well known fact that, if it hits the end-point 0 of the half-line at
some instant $t$, then for any instant $t' > t$ it will hit again 0
between $t$ and $t'$ an infinite number of times almost surely, but
still the cumulated time spent at zero is zero almost surely.  Slowly
reflected Brownian motion returns also almost surely an infinite
number of times after the first hit of the boundary, but the cumulated
time spent at zero (technically speaking the Lebesgue measure of those
instants in $[t,t']$ when the particle is at zero) will be
proportional to the time spent in an infinitesimal neighborhood
$(0,\dd x)$ of 0 (not containing 0!), the proportionality constant
being $\lambda / \dd x$.  Remarkably, however, for any $\lambda <
\infty$, the trajectory will almost surely never remain at $0$ for an
uninterrupted strictly positive time-interval. The two extreme values
of $\lambda$ are $\lambda = 0$ and $\lambda=\infty$. In the first case
zero time is spent at zero and, in fact, one recovers the ordinary
(instantaneous) reflection.  The second case with infinite time spent
at zero corresponds to absorption (more precisely, adsorption) at
zero.  Transposed to the case of Lagrangian dispersion, this means
that particles upon hitting each other either have no interaction
(case $\lambda = 0$), or they will tend to stay together for some
positive amount of cumulated time (case $0 < \lambda < 0$), or they
stick together permanently (case $\lambda = \infty$).
\vskip 0.15cm
 
We shall give strong general arguments in favor of the convergence of
the particle dispersion to the sticky process and shall prove the
convergence for specific quantities.  To analyze the effect of
stickiness on the passive transport, we shall construct and analyze
the behavior of the transition probabilities of the sticky dispersion
process that govern, in the appropriate $l_\nu,l_\kappa \rightarrow
0$ limit, the evolution of the 2-point correlation function of
advected tracer in the homogeneous and isotropic situation.  It
appears that the presence of stickiness induces anomalous scaling of
the stationary 2-point structure function of the forced tracer and
that it influences the rate of tracer energy condensation in the
constant mode.
For the two extreme cases of instantaneous reflection and absorption,
not only the particle dispersion process but the entire family of the
consistent $N$-particle Markov processes has been constructed in
\cite{LeJR1,LeJR2}.  In the sticky case, the existence and the
uniqueness of such a construction is still an open problem and is not
discussed further here.  Let us only note that ref.\ \cite{LeJR3} may
be interpreted as providing such a construction for the limiting case
of the one-dimensional flow with $\xi=0$.
\vskip 0.3cm

Let us recapitulate the situation that we are dealing with.  We are
studying the intermediate compressibility regime of the Kraichnan
model of passive advection with velocity field smoothed at scales
smaller than $l_\nu$ and with molecular diffusivity $\kappa$ dominant
on scales smaller than $l_\kappa$.  We shall be looking at the 2-point
function of the advected tracer in the isotropic sector or,
alternatively, at the PDF of the distance between two noisy Lagrangian
trajectories.  We want to know how those objects evolve in the limit
$l_\nu,l_\kappa \rightarrow 0$.  The remainder of the paper is
organized as follows.  In \secref{sect:stoch-proc} we shall discuss
the probabilistic aspects of the problem studying the one-dimensional
diffusion process describing the inter-particle distance.  The
analytic approach based on the spectral analysis of the generator of
the process will be developed in \secref{sect:linop}.  The analysis of
the limits $l_\nu,l_\kappa \rightarrow 0$ in those sections is based
on approximate calculations.  In \secref{sect:sticky-pr}, we study the
sticky limiting behavior of the trajectories obtained this way in more
detail and we analyze its implications for the tracer transport.  To
confirm the approximate analysis of the first sections, we rigorously
show in \secref{sect:proof} that, in the intermediate compressibility
regime, the stationary 2-point structure function of the forced tracer
indeed converges when $l_\nu,l_\kappa \to 0$ in the fine-tuned way to
the one corresponding to the sticky behavior of Lagrangian
trajectories.  Finally, in Conclusions, we summarize the obtained
results and mention some open problems that they raise.  Appendices
contain more technical material relevant for the main text.
\vskip 0.15cm

\subsection*{Acknowledgements}

The authors thank for hospitality the Erwin Schr\"{o}dinger Institute
in Vienna and the Institute for Advanced Study at Princeton where
parts of this work were done.  They are grateful to Y.~Le~Jan and
O.~Raimond for numerous explanations and discussions and to W.~E 
and E.~Vanden-Eijnden for helpful exchanges. K.G.\ acknowledges
the support of the von Neumann Fund during his stay at the IAS.

\section{The stochastic process viewpoint}
\label{sect:stoch-proc}

\subsection{Natural scale and speed measure}

In the intermediate compressibility regime, zero is a regular boundary
for the unregularized operator $M$ of Eq.~\eqref{eq:M} according to
the Feller criteria.  That is, if $M$ is viewed as the generator of a
stochastic process $r(t)$ on $\RR_+$ then the realizations of the
process may hit zero and come back from it, with positive probability.
In order to completely describe the process, one needs to specify a
boundary condition for $M$ at zero.  On the other hand, for
$l_\nu,l_\kappa$ finite, the regularized generator $M_{\nu,\kappa}$ of
Eq.~\eqref{eq:riop} should be taken with Neumann boundary condition
at zero.  As we have discussed above, this is because the problem on
the half-line arose as the rotationally invariant sector of a
non-degenerate problem defined on the $d$-dimensional space.
\vskip 0.15cm

To see what boundary condition is obtained when $l_\nu,l_\kappa
\rightarrow 0$, it is useful to study the natural scale and the speed
measure (see \appref{apx:1d-dif}) of the process $r(t)$ with positive
$l_\nu,l_\kappa$ and to determine their behavior in the aforementioned
limit.  The natural scale $S(r)$ is the (positively oriented)
coordinate in which the generator of the diffusion $r(t)$ is without
drift so that $S(r(t))$ is a martingale (there is of course an
equivalence class of such coordinates, related by affine
transformations).  $S(r)$ is strictly increasing and continuous in
$r$, so that the function $s([r_1,r_2]) \equiv S(r_2) - S(r_1)$
defined on intervals can be extended to a measure on $\RR_+$
absolutely continuous w.r.t.\,\,the Lebesgue measure).  The measure
$\dd s$ is sometimes called the natural scale measure. Often it is
more practical to work with the density of a measure than the measure
itself.  For a measure $\dd\mu$ and a coordinate $r$ on $\RR_+$ we
denote by $\mu(r)$ the density of $\dd\mu$ with respect to $r$.
In the case of $\dd s$ we have in particular $s(r) =\frac{\dd S(r)}{\dd r}$.
\vskip 0.15cm

In terms of the functions $f_1(r)$, $f_2(r)$ entering the generator
$M_{\nu,\kappa}$ one can choose for the density with respect to $r$ of
the natural scale measure
\begin{equation}
\label{eq:scale-meas}
  s(r)
=
  \exp \left(
    -\int_{r_0}^r
       \frac{f_2(r')}
            {f_1(r')}
     \,\dd r'
  \right)
\end{equation}
where $r_0 > 0$ is an arbitrary point of the open half-line $\RR_+^*$.
It is easily checked that in the corresponding coordinate $S$, the
process becomes driftless (no first order derivative):
\begin{equation}
  M_{\nu,\kappa}
= 
  {\tilde D}_0\,f_1(r)\,s(r)^2 \partial_{S}^2
\ \equiv\ 
  \tilde D_0 \, m(S)^{-1}\partial_S^2
\,.
\nonumber
\end{equation}
\vskip 0.1cm

We shall define the speed measure as the measure $\dd m$ with density
$m(S)$ with respect to the coordinate $S$. For convenience, we have
multiplied the speed measure by $2\tilde D_0$ with respect to the
conventions of \appref{apx:1d-dif} in order to avoid such factors in
later expressions.  The diffusion process $S(t) = S(r(t))$ is then a
(random, i.e.\ realization dependent) re-parametrization of the
Brownian motion $W(\tau)$, i.e.\ $S(t) = W(\tau(t))$, with
$$
  2\tilde D_0\, \frac{\dd t}{\dd\tau}
\,=\,
  m(S)|_{S=W(\tau)}
\,.
$$
For later use we also calculate the density of the speed measure with
respect to the coordinate $r$:
\begin{equation}
\label{eq:speed-meas}
  m(r)
=
  m(S) \frac{\dd S}{\dd r}
=
  [f_1(r) s(r)]^{-1}
=
  \frac{1}
       {f_1(r)}
  \exp \left(
    \int_{r_0}^r
      \frac{f_2(r')}
           {f_1(r')}
    \,\dd r'
  \right)
\end{equation}
with the same $r_0$ as for the natural scale above.
\vskip 0.15cm

We expect that the process obtained in the $\,l_\kappa,l_\nu
\rightarrow 0\,$ limit corresponds to the natural scale and the speed
measure that are appropriate limits of the same objects for
$l_\kappa,l_\nu$ finite.  In the next subsections, we shall analyze
those limits in an approximate way.

\subsection{Approximate calculations}

We should know the dependence of functions $f_1$ and $f_2$ on the
regularization scales $l_\nu,l_\kappa$.  If $l_\nu < l_\kappa$ then
the smoothing of the velocity field is only significant at scales
where it is already not the advection but the diffusion term that
dominates, that is to say, the smoothing of the velocity plays
no role and we may take $l_\nu = 0$ immediately.  This case goes
without difficulties and has been studied in \cite{G}, see also
\cite{EVDE1}.  The case which is interesting for us is the other one,
when $l_\kappa < l_\nu$.  In this case we feel the diffusivity between
0 and $l_\kappa$, then we feel the smoothing of the velocity field
between $l_\kappa$ and $l_\nu$ and finally above $l_\nu$ we are in the
genuine Kraichnan regime with some scaling exponent $\xi$ of the
second order velocity structure function.
\vskip 0.15cm

In the present section, the calculations will be made by replacing
functions $f_1,f_2$ by functions glued piecewise from pure powers,
representing the different scaling behaviors in different
subintervals.  The gluing is done so that the function $f_1$ stays
continuous.  That the model functions obtained this way are indeed
correct approximations of functions $f_1,f_2$ for the regularized
Kraichnan model is shown in \appref{apx:f1-f2}.  Thus for $r > l_\nu$
we shall take $f_1(r)$ and $f_2(r)$ as in the (scale-invariant)
Kraichnan model, that is $f_1(r) = r^\xi$ and $f_2(r) =
a_\xi r^{-1} f_1(r)$ with $a_\xi$ defined in \eqref{def:a_xi}.  For
$l_\kappa < r < l_\nu$ the functions $f_1,f_2$ will behave as in the
smooth Kraichnan model (i.e.\ for $\xi = 2$), and we have to match
them at $l_\nu$ with the values already given above (approximately;
what matters, as we shall see, is only the order of magnitude and the
ratio of the functions).  Thus we shall take $f_1(r) = r^2
l_\nu^{\xi-2}$ and $f_2(r) = a_2 r^{-1} f_1(r)$.  In the same way
for $r < l_\kappa$ the functions $f_1$ and $f_2$ will be like in the
pure diffusive case which, incidentally, is the same as the Kraichnan
model with $\xi=0$.  With the same kind of matching as above, but now
at $l_\kappa$, we get $f_1(r) = l_\kappa^2 l_\nu^{\xi-2}$ and $f_2(r)
= (d-1)r^{-1} f_1(r)$ there.  \tabref{tab:f1-f2} sums up our choices
for the approximate versions of $f_1$ and $f_2$.
\begin{table}[h]
\begin{center}
\vskip 0.3cm
\begin{tabular}{|l|c|c|}
\hline
  \multicolumn{1}{|c|}{Scale} & $f_1(r)$ & $r f_2(r)/f_1(r)$ \\
\hline \hline
  \rule[-1ex]{0mm}{3.2ex}\!
  $r \in [0,l_\kappa]$ & $l_\kappa^2 l_\nu^{\xi-2}$ & $d-1$ \\
\hline
  \rule[-1ex]{0mm}{3.2ex}\!
  $r \in [l_\kappa,l_\nu]$ & $r^2 l_\nu^{\xi-2}$ & $a_2$ \\
\hline
  \rule[-1ex]{0mm}{3.2ex}\!
  $r \in [l_\nu,\infty]$ & $r^\xi$ & $a_\xi$ \\
\hline
\end{tabular}
\caption{Approximate versions of functions $f_1$ and $f_2$}
\label{tab:f1-f2}
\end{center}
\vskip -0.3cm
\end{table}

With the pure power choices for $f_1$ and $f_2$, it is straightforward 
to calculate the natural scale and the speed
measure on the intervals $[0,l_\kappa],[l_\kappa,l_\nu]$ and $[l_\nu,\infty)$.
Let us first evaluate the ubiquitous subexpression $s(r) =
\exp(-\int_{r_0}^r f_2(r')/f_1(r') \,\dd r')$.  We see that in each of
the above intervals, $f_2(r')/f_1(r') = a/r'$ with some $a$ constant
on the interval, so the integral evaluates to logarithms.  Again,
results are summed up in \tabref{tab:s,m}.
\begin{table}[h]
\begin{center}
\vskip 0.3cm
\begin{tabular}{|l|c|c|}
\hline
  \multicolumn{1}{|c|}{Scale} & $s(r)$ & $m(r)$ \\
\hline \hline
  \rule[-1.2ex]{0mm}{3.7ex}\!
  $r \in [0,l_\kappa]$ &
  $(r_0/l_\nu)^{a_\xi} (l_\nu/l_\kappa)^{a_2} (l_\kappa/r)^{d-1}$ &
  $r_0^{-\xi}
   (l_\nu/r_0)^{a_\xi-\xi} (l_\kappa/l_\nu)^{a_2-2} (r/l_\kappa)^{d-1}$ \\
\hline
  \rule[-1.2ex]{0mm}{3.7ex}\!
  $r \in [l_\kappa,l_\nu]$ &
  $(r_0/l_\nu)^{a_\xi} (l_\nu/r)^{a_2}$ &
  $r_0^{-\xi} (l_\nu/r_0)^{a_\xi-\xi} (r/l_\nu)^{a_2-2}$ \\
\hline
  \rule[-1.2ex]{0mm}{3.7ex}\!
  $r \in [l_\nu,\infty]$ &
  $(r_0/r)^{a_\xi}$ &
  $r_0^{-\xi} (r/r_0)^{a_\xi-\xi}$ \\
\hline
\end{tabular}
\caption{Densities of the natural scale and speed measures}
\label{tab:s,m}
\end{center}
\vskip -0.65cm
\end{table}

\subsection{Limit of the natural scale}

It is easy to see that if we maintain $r_0 > 0$ fixed, then the
density $s(r)$ of the natural scale measure tends pointwise to
\begin{equation}
\label{def:s_0}
  s_0(r)
\equiv
  r_0^{a_\xi} r^{-a_\xi}
\,.
\end{equation}
Recall that $s(r) = \frac{\dd S(r)}{\dd r}$.  Thus $S(r)$ has to be
the integral of $s(r)$, but we have the freedom to choose the constant
of integration.  We adopt the choice
$$
  S(r)
\equiv
  \int_0^{r_0} s_0(r') \,\dd r' +
  \int_{r_0}^r s(r')   \,\dd r'
\,.
$$
This definition has the advantage that, for $r > l_\nu$, the value of
$S(r)$ is independent of $l_\kappa,l_\nu$, because $s(r) = s_0(r)$
there.  In fact we have
\begin{equation}
\label{def:S_0}
  S(r)
=
  S_0(r)
\equiv
  \int_0^r s_0(r') \,\dd r'
=
  \frac{1}{1-a_\xi} r_0^{a_\xi} r^{1-a_\xi}
\qquad
  \text{if }\ r \geq l_\nu
\,.
\end{equation}
In particular, in the limit $l_\kappa,l_\nu \rightarrow 0$, the
natural scale function becomes equal to $S_0(r)$ for $r>0$.  Note that
$S_0(0) = 0$, observe, however, that if $d \geq 2$ then $S(0) =
-\infty$ as long as $l_\kappa$ is positive since in that case the
integral $\int_{l_\kappa}^{r} s(r') \,\dd r'$ diverges as $r$ goes to
zero.  Relation \eqref{def:S_0} may be inverted to give
\begin{equation}
\label{expr:r(S)}
  r(S)
=
  [(1-a_\xi) r_0^{-a_\xi} S]^{\frac{1}{1-a_\xi}}
\geq
  l_\nu
\qquad
  \text{if} \quad
  S \geq \frac{1}{1-a_\xi} r_0^{a_\xi} l_\nu^{1-a_\xi}
\,.
\end{equation}

\subsection{Limit of the speed measure}

Let us now calculate the speed measure of each of the intervals
$[0,l_\kappa]$, $[l_\kappa,l_\nu]$ and for the sake of completeness
$[l_\nu,R]$ for arbitrary $R > l_\nu$.
\vskip 0.15cm

For $[0,l_\kappa]$ we get
\begin{align*}
  m([0,l_\kappa])
\ &=\ 
  \int_0^{l_\kappa} m(r) \,\dd r
\ =\ 
  \int_0^{l_\kappa}
    \frac{1}{l_\kappa^2 l_\nu^{\xi-2}}
    \left(\frac{l_\nu}{r_0}\right)^{a_\xi}
    \left(\frac{l_\kappa}{l_\nu}\right)^{a_2}
    \left(\frac{r}{l_\kappa}\right)^{d-1}
  \,\dd r
\\
&=\ 
  \frac{r_0^{-a_\xi}}{d}
  l_\nu^{(a_\xi+1-\xi)+(1-a_2)}
  l_\kappa^{a_2-1}
\,.
\end{align*}
In the intermediate compressibility regime, $a_\xi + 1 - \xi > 0$ and
$1 - a_2 > 0$,
hence we have a positive power of $l_\nu$ and a negative power of
$\,l_\kappa$.  Depending on the way in which $l_\nu,l_\kappa$ go to
zero, $m([0,l_\kappa])$ can tend to zero, infinity or some fixed
constant.  For the finite limit one should have
$\,l_\nu^{(a_\xi+1-\xi) + (1-a_2)} \propto l_\kappa^{1-a_2}$, \,i.e.
\begin{equation}
\label{eq:prop}
  l_\kappa
\,\propto\,
  l_\nu^{1+\frac{a_\xi+1-\xi}{1-a_2}}
\end{equation}
The exponent of $\,l_\nu$ being greater than 1, this relation is
compatible with the condition $l_\kappa \ll l_\nu$ in the limit
$l_\nu,l_\kappa \rightarrow 0$.  In a moment we shall explain how the
limit $l_\nu,l_\kappa \rightarrow 0$ taken with condition
\eqref{eq:prop} corresponds to the sticky boundary condition.  For now
let us simply give some equivalent formulations of \eqref{eq:prop}.
From \eqref{eq:f_1}, it can be seen that $2 \kappa = \tilde{D}_0
f_1(0)$.  Within our current modeling of $f_1$, we have $f_1(0) =
l_\nu^{\xi-2} l_\kappa^2$ so that
\begin{equation}
\label{expr:kappa}
  \kappa
=
  \frac{_1}{^2}
  \tilde{D}_0\, l_\nu^{\xi-2}\ l_\kappa^2\,.
\end{equation}
We shall use this equation to fix the relation between $\kappa$ and
$l_\kappa$ also in the exact approach.  Recalling definition
\eqref{eq:visc}, we see that the Prandtl number $\mathit{Pr} \equiv
\frac{\nu}{\kappa}$ is proportional to $(l_\nu/l_\kappa)^2$ for
$l_\kappa\ll l_\nu$ so that imposing relation \eqref{eq:prop} is
equivalent to
$$
  \kappa
\,\propto\,
l_\nu^{\xi + 2\frac{a_\xi+1-\xi}{1-a_2}}
\qquad \text{or} \qquad
  \mathit{Pr}
\,\propto\,
l_\nu^{-2\frac{a_\xi+1-\xi}{1-a_2}}\,.
$$
In particular $\kappa$ goes to zero while $\mathit{Pr}$ goes to
infinity when $l_\nu$ goes to zero.
\vskip 0.15cm

The calculation of the speed measure of the interval $[l_\kappa,l_\nu]$ is
performed similarly:
\begin{align*}
  m([l_\kappa,l_\nu])
\ &=\ 
  \int_{l_\kappa}^{l_\nu} m(r) \,\dd r
\ =\ 
  \int_{l_\kappa}^{l_\nu}
    \frac{1}{r^2 l_\nu^{\xi-2}}
    \left(\frac{l_\nu}{r_0}\right)^{a_\xi}
    \left(\frac{r}{l_\nu}\right)^{a_2}
  \,\dd r
\\
&=\ 
    \frac{r_0^{-a_\xi}}{a_2 - 1}
  l_\nu^{a_\xi - a_2 + 2 - \xi}
  \left(l_\nu^{a_2-1} - l_\kappa^{a_2-1}\right).
\end{align*}
Now, since in the intermediate compressibility regime $\,a_2 - 1 < 0$,
in the limit $l_\nu,l_\kappa \rightarrow 0$ it will be
$l_\kappa^{a_2-1}$ that dominates $l_\nu^{a_2-1}$ so that
$$
  m([l_\kappa,l_\nu])
\,\approx\,
  \frac{r_0^{-a_\xi}}{1 - a_2}
  l_\nu^{(a_\xi+1-\xi) + (1-a_2)}
  l_\kappa^{a_2-1}.
$$
We are in exactly the same situation as above and the same conclusions
hold.
\vskip 0.15cm

Finally the speed measure of the interval $[l_\nu,R]$ for some
arbitrary $R>l_\nu$ is
\begin{align*}
  m([l_\nu,R])
\ &=\ 
  \int_{l_\nu}^{R} m(r) \,\dd r
\ =\ 
  \int_{l_\nu}^{R}
    r^{-\xi}
    \left(\frac{r}{r_0}\right)^{a_\xi}
  \,\dd r
\\
&=\ 
  \frac{r_0^{-a_\xi}}{a_\xi + 1 - \xi}
  \left(R^{a_\xi+1-\xi} - l_\nu^{a_\xi+1-\xi}\right)
\,\approx\,
  \frac{r_0^{-a_\xi}}{a_\xi + 1 - \xi}
  R^{a_\xi+1-\xi}
\end{align*}
and it tends to a finite limit when $l_\nu,l_\kappa \rightarrow
0$.
\vskip 0.15cm

At this point we may describe explicitly the speed measure $\dd m_0$
obtained as the weak limit of speed measures for positive
$l_\nu,l_\kappa$ when $l_\nu,l_\kappa\rightarrow 0$ in such a way that
\begin{equation}
\label{def:lambda}
  l_\nu^{(a_\xi+1-\xi)+(1-a_2)} 
  l_\kappa^{a_2-1}
\,\to\,\,
  \lambda
\end{equation}
for some $\lambda \in [0,+\infty]$ (note that $\lambda$ has dimension
$\,{(\dimface{length})}^{a_\xi+1-\xi}$).  On the open half-line of
$r>0$ the density of $\,\dd m_0$ is
\begin{equation}
\label{expr:m-dens-stoch}
  m_0(r)
=
  r_0^{-a_\xi} r^{a_\xi - \xi}.
\end{equation}
Besides, $\dd m_0$ has a mass at zero, given by
\begin{equation}
\label{expr:m-mass-stoch}
  m_0(\{0\})
=
  \lim_{l_\nu,l_\kappa \rightarrow 0}
    (m([0,l_\kappa]) + m([l_\kappa,l_\nu]))
\,=\,
  \frac{d+1-a_2}{(1-a_2)d} \,
  r_0^{-a_\xi} \, \lambda\,.
\end{equation}
It is convenient to characterize the ``stickiness'' of the boundary at
$r=0$ by the quantity $\tilde{\lambda} \equiv m_0(\{0\}) /
[r^{\xi-a_\xi} m_0(r)]_{r=0}$ which we shall call
the {\bf glue parameter}.  We obtain
\begin{equation}
\label{expr:glue-param-1}
  \tilde\lambda
=
  \frac{d+1-a_2}{(1-a_2)d} \lambda
\end{equation}
This relation will be compared to the value produced by another
approach in the next section, and finally to the exact result
\eqref{expr:glue-param-3} calculated using the precise forms of
$f_1,f_2$ instead of the approximate versions of \tabref{tab:f1-f2}.

\subsection{Convergence on the natural scale}

Here we shall show that the operations of changing coordinates from
the original scale $r$ to natural scale $S$ and of taking the weak
limit of the speed measure $\dd m$ when $l_\nu,l_\kappa \to 0$
commute.  Observe that for $d=1$ the natural scale coincides with the
original one so we shall be preoccupied here only by the case $d \geq
2$.

In the preceding subsection we computed the limit of $m(r)$, working
on the original scale $r$.  Here we shall pass first to the natural
scale $S$ and then take the limit of $m(S)$ and finally check that the
result agrees with that of the preceding subsection.  This is in order
to exclude that we are in the kind of pathological case presented in
\appref{apx:patho}.  A priori such a situation could arise here
because the notions of weak convergence on the original scale and on
the natural scale do not coincide.
\vskip 0.15cm

The density of the speed measure w.r.t.\ the natural scale $S$ is
$m(S) = m(r) s^{-1}(r)$.  From \eqref{expr:r(S)} it is clear that for
any $S > 0$ we can take $l_\nu$ sufficiently small to have $r(S) >
l_\nu$ and thus $m(r) = m_0(r)$ and $s(r) = s_0(r)$ so that $m(S) =
m_0(r) s_0^{-1}(r) = m_0(S_0)$.  Hence the trivial convergence of the
density of the speed measure on the open half-line of positive $S$ to
$m_0(S_0)$ in agreement with the result of the preceding subsection.
\vskip 0.15cm

We still have to see what happens for $S < 0$ and to show that the
limit $m_0(S_0)$ of the density of the speed measure has a Dirac 
delta term at $\,S_0 = 0$.  
Let us evaluate for $S<0$ the asymptotic behavior of $r(S)$ as
$l_\kappa,l_\nu \rightarrow 0$.  Observe that $S(l_\nu) =
\int_0^{l_\nu} s_0(r') \,\dd r' > 0$.  So if $S<0$ then $r(S) <
l_\nu$.  Now
\begin{align*}
  S(l_\kappa)
&=
  \int_0^{l_\nu} s_0(r') \,\dd r' -
  \int_{l_\kappa}^{l_\nu} s(r') \,\dd r'
\\
&=
  \frac{1}{1-a_\xi} r_0^{a_\xi} l_\nu^{1-a_\xi} -
  \frac{1}{1-a_2  } r_0^{a_\xi} l_\nu^{1-a_\xi}
    \left[
      1 - \left(\frac{l_\kappa}{l_\nu}\right)^{1-a_2}
    \right]
\ \longrightarrow\ 
  0
\end{align*}
when $l_\nu\to0$ with $l_\kappa<l_\nu$ since both $1-a_\xi$ and
$1-a_2$ are positive.  This shows that for $S<0$ fixed, we must have
$r(S) < l_\kappa$ asymptotically.  Let us explicitly treat the case $d
> 2$.  The case $d=2$ is only a bit different because of the
logarithmic divergences.  For $r < l_\kappa$,
\begin{align*}
  S(r)
&=
  S(l_\kappa) -
  \int_r^{l_\kappa} s(r') \,\dd r'
\\
&=
  S(l_\kappa) -
  \frac{1}{2-d}\, r_0^{a_\xi}\, l_\nu^{a_2-a_\xi}\, l_\kappa^{d-1-a_2}
  \,(l_\kappa^{2-d} - r^{2-d})\,.
\end{align*}
Suppose that $r = o(l_\kappa)$ (consistency will be checked) so that
$l_\kappa^{2-d} = o(r^{2-d})$.  Recalling that $S(l_\kappa)
\rightarrow 0$, we infer that, for fixed $S<0$,
\begin{align*}
  r(S)
&\sim
  \left[
    (2-d)\, r_0^{-a_\xi}\, l_\nu^{a_\xi-a_2} l_\kappa^{1-d+a_2} S
  \right]^{\frac{1}{2-d}}
\\
&=
  \left[
    (d-2)\, r_0^{-a_\xi} (-S)
  \right]^{-\frac{1}{d-2}}
  \left[
    l_\nu^{1-a_\xi} (l_\kappa/l_\nu)^{1-a_2}
  \right]^{\frac{1}{d-2}}
  l_\kappa
\,.
\end{align*}
We immediately check that $r(S) = o(l_\kappa)$ so that our estimates
are consistent, as promised.  It can also be seen that this
asymptotics for $r(S)$ is uniformly valid on $(-\infty,S]$ for any $S
< 0$.  Now we may write
\begin{align*}
  m(S)
&=
  [m(r) \, s^{-1}(r)]_{r=r(S)}
\\
&\sim
  \left[
    (d-2)^{-(d-1)} \, r_0^{a_\xi} \,
    (l_\kappa/l_\nu)^{1-a_2} \,
    l_\nu^{1-a_\xi}
  \right]^{\frac{2}{d-2}}
  l_\nu^{2-\xi} \,
  (-S)^{-[2 + \frac{2}{d-2}]}
\,.
\end{align*}
We infer that $m(S)$ decays as a power of $-S$ when $S \rightarrow
-\infty$, fast enough to be integrable.  We also see that the
coefficient in front of the power of $-S$ goes to zero when $l_\nu
\rightarrow 0$ with $l_\kappa<l_\nu$.
\vskip 0.15cm

We still have to show that $m_0(S)$ has a Dirac delta contribution
at zero.  This can now be seen because
$$
  \int_{-\infty}^{S(l_\nu)} m(S) \,\dd S
=
  \int_0^{l_\nu} m(r) \,\dd r
$$
and we have shown that the right hand side has the limit given by
\eqref{expr:m-mass-stoch}.  On the other hand we have just seen that
for any $S < 0$
$$
  \int_{-\infty}^S m(S') \,\dd S'
\ \longrightarrow\ 
  0
$$
and also that $S(l_\nu) > 0$ and $S(l_\nu) \rightarrow 0$.  This
permits to conclude that $m_0(S_0)$ has a Dirac delta contribution at
$S_0=0$ with the coefficient $m_0(\{0\})$.  An analogous reasoning for
the case $d=2$ gives the same result regarding the (weak) convergence
of the measure $\dd m$ on the natural scale.  The limit measure
coincides with the limiting speed measure in the $r$ coordinate
re-expressed in the limiting natural scale coordinate $S_0$.

\subsection{Conjecture}

It is legitimate to expect that the process $r(t)$ converges in an
appropriate sense to the one corresponding to the limit natural scale
function $S_0\geq0$ given in \eqref{def:S_0}, and the speed measure
with density
$$
  m_0(S_0)
=
  r_0^{-a_\xi/b_\xi} \,
  [(1-a_\xi) S_0]^{-2+\frac{1}{b_\xi}}
  \ +\
  m_0(\{0\}) \, \delta(S_0)
\,.
$$
Such a process is, in the natural scale coordinate $S_0$, a
reparametrization of the Brownian motion $|W(\tau)|$ instantaneously
reflecting at zero: $\,S_0(t)=|W(\tau(t))|$, see \appref{apx:1d-dif}.
Away from zero, $\,2\tilde D_0\,\frac{\dd t}{\dd \tau} =
m_0(S_0)|_{S_0=|W(\tau(t))|}$.  At zero, the reparametrized process
spends time proportional to $m_0(\{0\})$ times the local time at zero
of $|W(\tau)|$ (although never an uninterrupted open interval of
time). This is the ``sticky'' or ``slowly reflecting'' boundary
behavior \cite{Breiman,RW} with the glue parameter $\,\tilde\lambda
\propto m_0(\{0\})$, see \eqref{expr:glue-param-1}. The extreme cases
$\,\tilde\lambda=0\,$ and $\,\tilde\lambda=\infty\,$ correspond,
respectively, to instantaneous reflection and total absorption of the
process at zero.
\vskip 0.15cm

This was our first argument for the sticky boundary behavior.

\section{The linear operator viewpoint}
\label{sect:linop}

Our next argument is based on the study of the (generalized)
eigenfunctions of the operator $M_{\nu,\kappa}$ of Eq.~\eqref{eq:riop}
in the limit $l_\nu,l_\kappa \rightarrow 0$.  If we take $f_1,f_2$
composed as above from pure powers, then we are able to calculate
exactly the eigenfunctions and trace their limiting behavior.  This
argument is somewhat more shaky because it needs some explicit, not
totally realistic (though representative), form of the functions $f_1$
and $f_2$ (before, such a representation was really required to hold
only approximately).

\subsection{Derivation of the boundary condition at zero}

Let us consider the differential operator $\tilde{D}_0^{-1}
M_{\nu,\kappa} = f_1(r) \partial_r^2 + f_2(r) \partial_r$ with the
approximate form of functions $f_1$ and $f_2$ given in \tabref{tab:f1-f2}.
Thus for $r > l_\nu$ we take $f_1(r) = r^\xi$ and $f_2(r) =
a_\xi r^{-1} f_1(r)$ so on the interval $[l_\nu,\infty]$ a pair of
linearly independent eigenfunctions corresponding to the eigenvalue
$-E$ of $\tilde{D}_0^{-1} M_{\nu,\kappa}$ is
$$
  \Phi_E^\pm (r)
=
  r^{\frac{1-a_\xi}{2}}
  J_{\pm b_\xi} (\sqrt{E} \,r^{\frac{2-\xi}{2}})
$$
where $b_\xi$ is given by \eqref{def:b_xi}
and $J_b$ is the Bessel function of the first kind of parameter $b$.  
Note that these eigenfunctions
are independent of $l_\nu, l_\kappa$.  When we impose at zero a boundary
condition\footnote{
        {\em Exactly one}, given as $L[\Phi_E] = 0$ where $L$ is an
        operator that can be expressed as the limit at zero of some
        finite linear combination of derivatives of order 0 or higher,
        with not necessarily constant coefficients.}%
\hspace{-0.4em},\hspace{0.2em} in general the eigenfunctions of the
operator with boundary condition will form a one dimensional linear
subspace of the two dimensional linear space spanned by $\Phi_E^\pm$.
That is, if the linear combination $\tilde{c}_E^+ \Phi_E^+ +
\tilde{c}_E^- \Phi_E^-$ is an eigenfunction verifying the boundary
condition at zero, then $\tilde{c}_E^+/\tilde{c}_E^-$ is fixed
(depending on $E$ and the boundary condition, of course).  In part we
shall proceed in the opposite way.  First we calculate
$\tilde{c}_E^+/\tilde{c}_E^-$ for $l_\nu,l_\kappa$ finite. Next we
deduce the effective boundary condition at zero that would give the
same quotient if $l_\nu,l_\kappa$ were zero.  Finally we look at the
limit of $\tilde{c}_E^+/\tilde{c}_E^-$ when $l_\kappa,l_\nu
\rightarrow 0$, and deduce the ``limit'' of the effective boundary
condition.
\vskip 0.15cm

First of all we give explicitly a generating pair of eigenfunctions of
$\tilde{D}_0^{-1} M_{\nu,\kappa}$ associated with the eigenvalue $-E$,
on each of the intervals $[0,l_\kappa], [l_\kappa,l_\nu],
[l_\nu,\infty]$.  We shall denote these six functions $g_1^\pm,
g_2^\pm, g_3^\pm$ respectively.  Because at zero we have the Neumann
boundary condition, on $[0,l_\kappa]$ we give only the corresponding
eigenfunction:
$$
  g_1^- (r)
=
  C_1
  r^{\frac{2-d}{2}}
  J_{-\frac{2-d}{2}} (\sqrt{E/\kappa\,} \, r)
\quad \text{with} \quad
  C_1
=
  2^{-\frac{2-d}{2}}
  \Gamma \left( 1-\frac{2-d}{2} \right)
  (E/\kappa)^{\frac{2-d}{4}}
$$
(the normalization was chosen for later convenience).  On
$[l_\kappa,l_\nu]$ we may set
$$
  g_2^\pm (r)
=
  r^{\gamma_\pm(E,l_\nu)}
\quad \text{where} \quad
  \gamma_\pm(E,l_\nu)
=
  \frac{1-a_2 \pm \sqrt{(1 - a_2)^2 -
                        4 l_\nu^{2-\xi} E
                       }
       }
       {2}
$$
(recall that $1-a_2 > 0$ in the intermediate compressibility regime).
Finally on $[l_\nu,\infty]$ we take
$$
  g_3^\pm (r)
=
  2^{\pm b_\xi}
  \Gamma(1 \pm b_\xi)
  E^{\mp \frac{b_\xi}{2}}
  \Phi_E^\pm(r)\,.
$$
\vskip 0.15cm

In order to construct the eigenfunction of $M_{\nu,\kappa}$ on the
whole half-line $[0,\infty]$, we have to find the correct linear
combinations of the generating pairs on each sub-interval by matching
the functions and their derivatives at each border point.  Formally,
if we have to match $c_i^+ g_i^+ + c_i^- g_i^-$ with $c_j^+ g_j^+ +
c_j^- g_j^-$ at $r$, then, in matrix notation, we should have
$$
  \begin{pmatrix}
    g_i^+ (r) & g_i^- (r) \\
    {g_i^+}'(r) & {g_i^-}'(r)
  \end{pmatrix}
  \begin{pmatrix}
    c_i^+ \\
    c_i^-
  \end{pmatrix}
=
  \begin{pmatrix}
    g_j^+ (r) & g_j^- (r) \\
    {g_j^+}'(r) & {g_j^-}'(r)
  \end{pmatrix}
  \begin{pmatrix}
    c_j^+ \\
    c_j^-
  \end{pmatrix}
$$
where $g'(r) \equiv \partial_r g (r)$.  That is to say,
$$
  \begin{pmatrix}
    c_j^+ \\
    c_j^-
  \end{pmatrix}
=
  \begin{pmatrix}
    g_j^+ (r) & g_j^- (r) \\
    {g_j^+}'(r) & {g_j^-}'(r)
  \end{pmatrix}^{-1}
  \begin{pmatrix}
    g_i^+ (r) & g_i^- (r) \\
    {g_i^+}'(r) & {g_i^-}'(r)
  \end{pmatrix}
  \begin{pmatrix}
    c_i^+ \\
    c_i^-
  \end{pmatrix}.
$$
Now, because on $[0,l_\kappa]$ we take $g_1^-$ as the eigenfunction,
it follows that, by making the above described matchings, we get on
$[l_\nu,\infty]$ the linear combination $c_E^+ g_3^+ + c_E^- g_3^-$
with
\begin{multline}
\label{eq:chain}
\hspace{-1.35em}
  \begin{pmatrix}
    c_E^+ \\
    c_E^-
  \end{pmatrix}
\!=\!
  \begin{pmatrix}
     g_3^+   (l_\nu) &  g_3^-   (l_\nu) \\
    {g_3^+}' (l_\nu) & {g_3^-}' (l_\nu)
  \end{pmatrix}^{\!-1}\!\!
  \begin{pmatrix}
     g_2^+   (l_\nu) &  g_2^-   (l_\nu) \\
    {g_2^+}' (l_\nu) & {g_2^-}' (l_\nu)
  \end{pmatrix}
  \begin{pmatrix}
     g_2^+   (l_\kappa) &  g_2^-   (l_\kappa) \\
    {g_2^+}' (l_\kappa) & {g_2^-}' (l_\kappa)
  \end{pmatrix}^{\!-1}\!\!
  \begin{pmatrix}
     g_1^-   (l_\kappa) \\
    {g_1^-}' (l_\kappa)
  \end{pmatrix}.\!
\end{multline}
Note that
$$
  \frac{\tilde{c}_E^+}
       {\tilde{c}_E^-}
=
  \frac{\Gamma(1+b_\xi)}
       {\Gamma(1-b_\xi)}
  \left(\frac{4}{E}\right)^{b_\xi}
  \frac{c_E^+}
       {c_E^-}\,.
$$
For a general matrix {\scriptsize $\begin{pmatrix} a & b \\ c & d
\end{pmatrix}$} its inverse is
$\frac{1}{ad-bc}${\scriptsize $\begin{pmatrix} d & -b \\ -c & a
\end{pmatrix}$}.
We use this formula to rewrite \eqref{eq:chain}, but since we are only
interested in the ratio $c_E^+ / c_E^-$, we get rid of the
$\frac{1}{ad-bc}$ factor.  Thus
\begin{multline}
\label{eq:chain2}
\hspace{0.8cm}
  \begin{pmatrix}
    c_E^+ \\
    c_E^-
  \end{pmatrix}
\ \propto
\ \begin{pmatrix}
     {g_3^-}' (l_\nu) &  -g_3^-   (l_\nu) \\
    -{g_3^+}' (l_\nu) &   g_3^+   (l_\nu)
  \end{pmatrix}\!
  \begin{pmatrix}
     g_2^+   (l_\nu) &  g_2^-   (l_\nu) \\
    {g_2^+}' (l_\nu) & {g_2^-}' (l_\nu)
  \end{pmatrix}\!\\
  \cdot\,\begin{pmatrix}
     {g_2^-}' (l_\kappa) &  -g_2^-   (l_\kappa) \\
    -{g_2^+}' (l_\kappa) &   g_2^+   (l_\kappa)
  \end{pmatrix}\!
  \begin{pmatrix}
     g_1^-   (l_\kappa) \\
    {g_1^-}' (l_\kappa)
  \end{pmatrix}.\hspace{0.8cm}
\end{multline}
This gives an explicit formula for $\,c_E^+ / c_E^-\,$ for
$\,l_\kappa, l_\nu$ finite.  In \appref{apx:exp} we show that if we
take $l_\nu,l_\kappa$ to zero in a way that the limit \eqref{def:lambda}
exists (notation $\,l_\nu,l_\kappa 
\mathop{\to}\limits^\lambda 0\,$)
\,then
\begin{equation}
\label{eq:quot-coeff}
  \lim_{l_\nu,l_\kappa\mathop{\to}\limits^\lambda 0}
   \ \frac{c_E^+}{c_E^-}
\,=\, 
  -
  \frac{d+1-a_2}
       {(1-a_2) (1-a_\xi) d}
  \,\lambda\,E\,.
\end{equation}

It is not difficult to identify the boundary condition implied by this
relation.  Recall that the eigenfunctions of $\tilde{D}_0^{-1}
M_{\nu,\kappa}$ associated to the eigenvalue $-E$ are of the form
$\Phi_E(r) = c_E^+ g_3^+(r) + c_E^- g_3^-(r)$ for $r > l_\nu$, (that
is for $r>0$ if $l_\nu,l_\kappa \rightarrow 0$).  Furthermore we see
that
\begin{alignat*}{2}
& g_3^-(0)
=
  1\,,
\quad&\quad
& r^{a_\xi} {g_3^-}' (r) |_{r=0}
=
  0\,,
\\
& g_3^+(0)
=
  0\,,
\qquad&\qquad
& r^{a_\xi} {g_3^+}' (r) |_{r=0}
=
  1-a_\xi\,.
\end{alignat*}
From this we conclude that $\Phi_E$ verifies
\begin{equation}
\label{expr:bound-pre}
  \Phi_E(0)
=
  c_E^-
\qquad\text{and}\qquad
  r^{a_\xi} \Phi_E'(r) |_{r=0}
=
  (1-a_\xi) c_E^+\,.
\end{equation}

\noindent For convenience let us denote $\tilde{D}_0 r^{a_\xi}
\partial_r\equiv F$ and, as before, $M_{0,0}\equiv M$.  Now
\eqref{expr:bound-pre} with \eqref{eq:quot-coeff} imply that
$$
   \tilde{D}_0^{-1} (F \Phi_E)(0)
=
  -\tilde\lambda
   \,
  E \,
  \Phi_E(0)
$$
where $\tilde\lambda$ is given by \eqref{expr:glue-param-1}.
Since $\Phi_E$ is an eigenvector of $\tilde{D}_0^{-1} M$ with
eigenvalue $-E$, we obtain
\begin{equation}
\label{expr:boundary-cond}
  (F \Phi_E)(0)
\,=\,
  \tilde\lambda
   \,
  (M \Phi_E)(0)\,.
\end{equation}
This last form may be read as an independent of $E$ boundary condition
that all $\Phi_E$ verify.  It becomes then {\em the} boundary
condition for $M$.  Note that the coefficient $\tilde\lambda$ is in
$[0,+\infty]$.  We introduce thus the operator $M_{\tilde\lambda}$
which is the version of $M$ with the domain
$\mathcal{D}(M_{\tilde\lambda})$ corresponding to the boundary
condition \eqref{expr:boundary-cond}.  In the next subsection we shall
show that $M_{\tilde\lambda}$ is the Kolmogorov backwards generator of
a sticky process whose glue parameter is equal to $\tilde\lambda$.

\subsection{Interpretation of the boundary condition}

To see that boundary condition \eqref{expr:boundary-cond} corresponds
to a slowly reflecting boundary, we may calculate the mass at zero of
the speed measure of the diffusion process generated by
$M_{\tilde\lambda}$.  The speed measure may be defined, up to a
multiplicative constant, as the measure $\dd m$ with respect to which the
generator $M_{\tilde\lambda}$ is self-adjoint \cite{RY}.  It is
straightforward to see that for $r>0$ we may take for the density of
the speed measure with respect to $r$ (recall notation of
\secref{sect:stoch-proc})
\begin{equation}
\label{expr:m-dens-diff}
  m(r)
=
  r^{a_\xi - \xi}\,.
\end{equation}
We have in particular the relation $m(r) M = \tilde{D}_0 \partial_r
\circ r^{a_\xi} \partial_r$.  Here, as in the following, if we write
$M$ instead of $M_{\tilde\lambda}$, we mean not the operator with a
specified domain but only the corresponding differential expression.
Let us denote the $L^2$ scalar product with respect to the measure $\dd m$
by $(.\,,.)_m$.  For $M_{\tilde\lambda}$ to be symmetric with respect
to $\dd m$, we need that $(g_1,M g_2)_m = (M g_1, g_2)_m$ for any 
$g_1,g_2 \in\mathcal{D}(M_{\tilde\lambda})$.  Using \eqref{expr:m-dens-diff} 
for the density of $\dd m$ on the open half-line $(0,\infty)$, and putting in
a separate term the effect of a possible mass of $\dd m$ at zero, this may be
written as
\begin{multline}
\label{eq:M-sym}
   g_1(0) (M g_2)(0)\,m(\{0\}) +
  \int_{0^+}^\infty g_1(r) (M g_2)(r)\,m(r) \,\dd r
=\\
   (M g_1)(0) g_2(0)\,m(\{0\}) +
  \int_{0^+}^\infty (M g_1)(r) g_2(r)\,m(r) \,\dd r
\end{multline}
where the limit $0^+$ in the integrals indicates that the integral is
over the open half-line $(0,\infty)$.  The integral on the left hand
side may be rewritten using integration by parts as
\begin{multline*}
\quad
  \int_{0^+}^\infty  g_1(r) (M g_2)(r)\,m(r)\,\dd r
=\\
  g_1(0) (F g_2)(0) -
  (F g_1)(0) g_2(0) +
  \int_{0^+}^\infty (M  g_1)(r) g_2(r) \,m(r)\,\dd r\,.
\quad
\end{multline*}
We then see that \eqref{eq:M-sym} holds if and only if
$$
  [ (M g_1)(0)\,m(\{0\}) - (F g_1)(0)] \,\, g_2(0)
\,\,=\,\,
  g_1(0) \,\, [ (M g_2)(0)\,m(\{0\}) -(F g_2)(0)]
$$
(strictly speaking the above expression is defined in the limit $r
\rightarrow 0$).  This condition may only be verified for all $g_1,g_2
\in \mathcal{D}(M_{\tilde\lambda})$ if, for all $g
\in\mathcal{D}(M_{\tilde\lambda})$, $([m(\{0\}) M - F] \, g)(0) = 0$.
Recalling the boundary condition \eqref{expr:boundary-cond}, this
amounts to the equality
\begin{equation}
\label{expr:m-mass-diff}
  m(\{0\})
=
  \tilde\lambda\,,
\end{equation}
We infer this way that the speed measure of $M_{\tilde\lambda}$ has a
finite mass at zero, which means that $M_{\tilde\lambda}$ describes a
process that is slowly reflecting at zero.  From
\eqref{expr:m-mass-diff} and \eqref{expr:m-dens-diff} we see that the
glue parameter of the slow reflection is $\tilde\lambda$.
\vskip 0.15cm

More generally, the speed measure found in this section is coherent
with the one we got from the stochastic process treatment.  Compare
\eqref{expr:m-dens-stoch}, \eqref{expr:m-mass-stoch} with
\eqref{expr:m-dens-diff}, \eqref{expr:m-mass-diff}, respectively, to
see that the two speed measures are proportional (the second equals
$r_0^{a_\xi}$ times the first one).

\section{The sticky process}
\label{sect:sticky-pr}

Now that we know the effective time evolution operator on large
scales, we may inquire how the corresponding transition probabilities
of the Lagrangian dispersion and other related quantities look like,
what is their long time behavior and the induced effect on tracer
transport.

\subsection{Transition probabilities}

The transition probabilities $P_{\tilde{\lambda}}^t(r;\dd r')$ for the
process with generator $M_{\tilde\lambda}$ are given by the
self-adjoint exponential operators $e^{t M_{\tilde\lambda}}$ defined
in $L^2(m(\dd r))$.  It will be, however, more convenient to use the
more standard $L^2(\dd r)$ conventions for the kernels and explicitly
write the contribution due to the atomic term of $\dd m$.  In such
notation,
\begin{equation}
\label{eq:trans-proba}
  P_{\tilde{\lambda}}^t(r;\dd r')
\ =\ 
  e^{tM_{\tilde\lambda}}(r,r') \,\dd r' \,+\,
  \tilde\lambda \,
  \left[
    e^{tM_{\tilde\lambda}}(r,\rho) \, \rho^{\xi-a_\xi}
  \right]_{\rho=0}
  \,\delta(r') \,\dd r'
\,.
\end{equation}
The term concentrated at zero assures the conservation of probability
$\int P_{\tilde{\lambda}}^t(r;\dd r')=1$ as is shown in
\appref{apx:cons-proba}.  Its presence implies that sticky
trajectories spend together positive time.  In particular, the mean of
the time that two trajectories starting at moment zero at distance $r$
spend together up to moment $t$ is $\,\tilde\lambda \int_0^t
[e^{sM_{\tilde\lambda}} (r,\rho) \, \rho^{\xi-a_\xi}]_{\rho=0} \,\dd
s$.
\vskip 0.15cm
  
To compute the kernel $\,e^{tM_{\tilde\lambda}}(r,r')$, it will be more
convenient to work with Schr\"{o}dinger type operators.  To this end,
we shall first change variables by setting $u = \frac{2}{2-\xi}
\tilde{D}_0^{-1/2} r^{\frac{2-\xi}{2}}$.  Recall from
\eqref{expr:gen-Bessel} that this transforms $M$ to the generator of a
Bessel process of parameter $-b$ with $b$ given by \eqref{def:b_xi}
(to ease the notation, we drop the subscript $\xi$).  After passing to
the new variable, we conjugate $M$ by the multiplication by $u^{b -
\frac{1}{2}}$ and finally multiply it by $-1$ to obtain the
differential operator $N$ of the form
$$
  N
\equiv
  -\partial_u^2 + \frac{b^2-\frac{1}{4}}{u^2}\,.
$$
The boundary condition \eqref{expr:boundary-cond} imposed on
$M_{\tilde\lambda}$ becomes for $N$ the condition
\begin{equation}
\label{expr:N-bc0}
  u^{1-2b} \partial_u u^{b-\frac{1}{2}} \varphi(u) |_{u=0}
=
  - \mu\,C_b^{-1}\,u^{b-\frac{1}{2}} N \varphi(u)|_{u=0}
\end{equation}
where $C_b = 2^{2b-1}\frac{\Gamma(b)}{\Gamma(1-b)}$ is introduced for
convenience, and
\begin{equation}
  \mu
\ =\ 
  \frac{(2-\xi)^{2b-1}\,\Gamma(b)}
       {\Gamma(1-b)}
  \,\tilde{D}_0^{b-1}
\,\tilde\lambda
\nonumber
\end{equation}
is a positive constant proportional to $\tilde\lambda$.  We shall
denote by $N_\mu$ the operator $N$ with this boundary condition.
Note for later use the relationship
\begin{equation}
\label{eq:M-N}
  e^{t M_{\tilde{\lambda}}} (r,r')
  \,\dd r'
=
  u(r)^{b-\frac{1}{2}}
  \, e^{-t N_\mu} (u(r),u(r'))
  \, u(r')^{\frac{1}{2}-b}
  \,\dd u(r')
\end{equation}
\vskip 0.15cm

The spectral properties of the operator $N_\mu$ are studied in detail
in \appref{apx:spec-meas-N}.  Here we rely on those results.  The
spectrum of $N_\mu$ is $\RR_+$.  The (generalized) eigenfunction
$\varphi_{\mu,E}(u)$ of this operator, associated to the eigenvalue $E
\geq 0$, may be taken as
\begin{equation}
\label{eq:eig-fct}
  \varphi_{\mu,E}(u)
=
  u^{\frac{1}{2}}
  \left[
    J_{-b}(\sqrt{E} u) -
    \mu E^{1-b} J_b(\sqrt{E} u)
  \right].
\end{equation}
The (scalar) spectral measure corresponding to this choice (of
normalization) of the eigenfunctions is
\begin{equation}
\label{eq:scal-spec-m}
  \dd\nu_\mu(E)
=
  \frac{\dd E}
       {2(1 - 2\mu E^{1-b} \cos(b\pi) + \mu^2 E^{2(1-b)})}\,.
\end{equation}
The kernel of the exponential of the operator $N_\mu$ may be written
using spectral calculus as
\begin{equation}
\label{expr:exp-N-spec}
  \exp(-t N_\mu)(u,v)
=
  \int_0^\infty
    e^{-tE}\, \varphi_{\mu,E}(u)\, \varphi_{\mu,E}(v)
  \,\dd\nu_\mu(E)\,.
\end{equation}
\vskip 0.1cm

An important aspect of the flow is the behavior of two particles
released at the same point at the same time.  The evolution of the
inter-particle distance of such a pair is described by the $r \to 0$
limit of $P_{\tilde{\lambda}}^t(r;\dd r')$.  Substituting
\eqref{expr:exp-N-spec} into \eqref{eq:M-N} and that in turn into
\eqref{eq:trans-proba}, we obtain:
\begin{equation}
\label{eq:sep-coal}
\begin{split}
  \lim_{r \to 0}\ P_{\tilde{\lambda}}^t(r;\dd r')
\ =\ 
& \frac{_{2^b}}{^{\Gamma(1-b)}}
  \Big(
    \int_0^\infty
      e^{-tE} \, E^{-\frac{b}{2}} \,
      \varphi_{\mu,E}(u(r'))
    \,\dd\nu_\mu(E)
  \Big)
  \,u(r')^{\frac{1}{2}-b}
  \,\dd u(r')\,
\\
& +\,\frac{_{2(2-\xi)^{2b-1}{\tilde D}_0^{b-1}}}
  {^{\Gamma(1-b)^2}} \, \tilde\lambda\,
  \Big(
    \int_0^\infty
      e^{-Et} \, E^{-b}
    \,\dd\nu_\mu(E)
  \Big)
  \,\delta(r') \,\dd r'\,.
\hspace{1em}
\end{split}
\end{equation}
The first term on the right hand side is the regular contribution
absolutely continuous w.r.t.\ $\dd r'$.  It describes the probability
that a pair of particles starting together at time zero are separated
at time $t$ by some finite distance $r'$ with $\dd r'$ precision.
Presence of such a term indicates that the sticky Lagrangian flow
should be stochastic in each velocity realization, just as the
instantaneously reflecting flow corresponding to $\tilde\lambda = 0$.
The second term on the right hand side of \eqref{eq:sep-coal} is
concentrated at $r'=0$ and describes the probability that two
trajectories starting together will be together at time $t$.  It is
absent for $\tilde\lambda=0$.  Recall, however, from \cite{GV} that in
the strongly compressible phase $\wp > \frac{d}{\xi^2}$ one has
$\lim\limits_{r \to 0} \,\, P_{\tilde{\lambda}}^t(r;\dd r') = \delta(r')
\,\dd r'$ signaling that the Lagrangian flow is deterministic there.
Appearance of both regular and singular contributions to
\eqref{eq:sep-coal} in the sticky flow is one of the indications of a
hybrid nature of this case.
\vskip 0.15cm

In the special instance of $\mu = 0$ or $\mu = \infty$, corresponding
to instantaneously reflecting or absorbing boundary at $r=0$, the
integral in \eqref{expr:exp-N-spec} may be calculated explicitly.  We
make the change of variables $E = z^2$ and then use Weber's formula,
Eq.~(6.633.2) of \cite{GR}, to obtain
\begin{align*}
  \exp(-t N_0)(u,v)
&=
  \int_0^\infty
    e^{-tE}
    u^{\frac{1}{2}} J_{-b}(\sqrt{E} u)
    v^{\frac{1}{2}} J_{-b}(\sqrt{E} v)
  \,\frac{\dd E}{2}
\\
&\hspace{-1.5cm}=
  u^{\frac{1}{2}} v^{\frac{1}{2}}
  \int_0^\infty
    e^{-tz^2}
    J_{-b}(z u)
    J_{-b}(z v)
  \, z \,\dd z
=
  \frac{\sqrt{uv}}{2t}
  \exp\left(-\frac{u^2+v^2}{4t}\right)
  I_{-b} \left(\frac{uv}{2t}\right)
\end{align*}
where $I_{-b}$ is the modified Bessel function of the first kind of
index $-b$. \,Analogously,
\begin{align*}
  \exp(-t N_{\infty})(u,v)
&=
  \int_0^\infty
    e^{-tE}
    u^{\frac{1}{2}} J_b(\sqrt{E} u)
    v^{\frac{1}{2}} J_b(\sqrt{E} v)
  \,\frac{\dd E}{2}
\\
&=
  \frac{\sqrt{uv}}{2t}
  \exp\left(-\frac{u^2+v^2}{4t}\right)
  I_b \left(\frac{uv}{2t}\right)\,.
\end{align*}
We then have
\begin{equation}
\label{expr:Bess-ccl}
  \exp(t M_{_0^\infty}) (r,r')
=
  \frac{_1}
       {^{\tilde{D}_0 (2-\xi) t}}\,r^\frac{1-a_\xi}{2}\,
  {r'}^\frac{a_\xi+1-2\xi}{2}
  \exp\left(
    -\frac{_{r^{2-\xi} + {r'}^{2-\xi}}}
          {^{\tilde{D}_0 (2-\xi)^2 t}}
  \right)
  I_{\pm b} \left(
    \frac{_{2 (r r')^\frac{2-\xi}{2}}}
         {^{\tilde{D}_0 (2-\xi)^2 t}}
  \right).
\end{equation}
It can be easily shown that these results are in agreement with the
well known transition probabilities of the Bessel process with
reflecting or absorbing boundary at zero \cite{HBM}.
\vskip 0.15cm

Some other quantities of interest are the hitting times $H_{r'}$
at some point $r'$, in particular, the doubling and halving times 
of the inter-particle distance, see e.g. \cite{CHGKV}, Sect.\,2C.  
Their expectations are given by the following formulae (with notations 
of \appref{apx:1d-dif})
$$
  \,\EE_r(e^{-\alpha H_{r'}})
=
  \frac{r^\frac{\xi-2a_\xi}{4}
        \phi_{\mu,-\alpha}(
          \frac{2 \tilde{D}_0^{-1/2}}{2-\xi} r^\frac{2-\xi}{2}
        )
       }
       {{r'}^\frac{\xi-2a_\xi}{4}
        \phi_{\mu,-\alpha}(
          \frac{2 \tilde{D}_0^{-1/2}}{2-\xi} {r'}^\frac{2-\xi}{2}
        )
       }
\qquad\text{if \ \,$r < r'$}
$$
and
$$
  \EE_r(e^{-\alpha H_{r'}})
=
  \frac{r^\frac{\xi-2a_\xi}{4}
        \psi_{-\alpha}(
          \frac{2 \tilde{D}_0^{-1/2}}{2-\xi} r^\frac{2-\xi}{2}
        )
       }
       {{r'}^\frac{\xi-2a_\xi}{4}
        \psi_{-\alpha}(
          \frac{2 \tilde{D}_0^{-1/2}}{2-\xi} {r'}^\frac{2-\xi}{2}
        )
       }
\qquad\,\,\text{if \ \,$r > r'$}
$$
where $\phi_{\mu,-\alpha}$ is the solution of the differential
equation $(N+\alpha)\phi = 0$ that verifies the boundary condition
\eqref{expr:N-bc0} at zero and $\psi_{-\alpha}$ is the solution that
tends to zero at infinity. We refer here to
\appref{apx:1d-dif} and \appref{apx:spec-meas-N}, in particular to
formula \eqref{expr:hitting} of the former and formulae
\eqref{expr:ef-N-bc0} and \eqref{expr:ef-N-bcinfty} of the latter.

\subsection{Long time asymptotics}

The sticky transition probability densities
$P_{\tilde{\lambda}}^t(r;\dd r')$ of the inter-particle distance
determine the behavior of the passive tracer 2-point function obtained
in the corresponding limit $l_\kappa,l_\nu \to 0$.  The free decay of
the tracer 2-point function is described directly by the transition
probability $P_{\tilde{\lambda}}^t(r;\dd r')$.  Indeed, for an initial
tracer distribution with a homogeneous isotropic 2-point correlation
function $F(0,r)$, at time $t$ the tracer distribution will be
\begin{equation}
\label{eq:2ptscr}
  F(t,r)
=
  \int_0^\infty F(0,r')\, P_{\tilde{\lambda}}^t(r;\dd r')
\,,
\end{equation}
as follows from \eqref{eq:2ptsc}.  Thus the long time decay of the
tracer 2-point function is determined by the large $t$ asymptotics of
$e^{t M_{\tilde\lambda}}(r,r')$.
Similarly, with isotropic forcing as in \eqref{eq:fc} and, for
simplicity, no tracer at time zero,
\begin{equation}
\label{eq:2ptscfr}
  F(t,r)
=
  \int_0^t \dd s
    \int_0^\infty \chi(r')\,P_{\tilde{\lambda}}^s(r;\dd r')
\,,
\end{equation} 
as follows from \eqref{eq:2ptscf}.  Thus the long time behavior of the
forced tracer 2-point function is determined by the large $t$
asymptotics of $\int_0^t e^{s M_{\tilde\lambda}}(r,r') \,\dd s$.
We shall assume fast decay for large $r$ of both $F(0,r)$ in the unforced
case and of $\chi(r)$ in the forced one.
\vskip 0.15cm

As before, it will be more convenient to study instead of
$M_{\tilde\lambda}$ the operator $N_\mu$.  To obtain the large $t$
behavior of $\,e^{-tN_\mu}(u,v)$ or of $L_\mu(t;u,v) \equiv \int_0^t
e^{-s N_\mu}(u,v) \,\dd s$, we may consider their Laplace transforms
given for $\alpha > 0$ (where $\alpha$ is the variable conjugate to
$t$) by the resolvent kernel
$$
  (N_\mu+\alpha)^{-1}(u,v)
\ =\
  \int_0^\infty e^{-\alpha t}\, e^{-tN_\mu}(u,v) \,\dd t
$$
and by $\,\frac{1}{\alpha}\,(N_\mu+\alpha)^{-1}(u,v)$, respectively.  
By the well
known Tauberian-Abelian theorem, for any real numbers $-1 < p_N <
\ldots < p_1 < \infty$, the following behaviors of the function $f(t)$
and of its Laplace transform $\hat f(\alpha)$ are equivalent
\begin{xxalignat}{3}
\ \quad&\mathrm{(i)}&
  f(t)
&=
  \sum_{i=1}^N c_i t^{p_i} + o(t^{p_N})
&\text{near $\,t = \infty$\,, }\qquad
\\
\ \quad&\mathrm{(ii)}&
  \hat{f}(\alpha)
&=
  \sum_{i=1}^N c_i \Gamma(p_i + 1) \alpha^{-p_i-1} +
o(\alpha^{-p_N-1})
&\text{near $\,\alpha = 0^+$ }
.\,\,\,\ \quad
\end{xxalignat}
The resolvent $(N_\mu + \alpha)^{-1}$ of $N_\mu$ is studied in
\appref{apx:spec-meas-N}.  It can be written like in \eqref{expr:G}
(except for the change of sign of $\alpha$) and expanded in $\alpha$
for $\alpha$ small to all orders $\leq 0$.  It is easy to see that,
for $\mu < \infty$, in the numerator of \eqref{expr:G} only the
leading term is of order strictly smaller than zero and there is
exactly one term of order zero.  More exactly, one obtains:
\begin{align*}
  (N_\mu+\alpha &)^{-1}(u,v)
=
  G_{\mu,-\alpha}(u,v)
\\
&=
  C_b\,
  (uv)^{\frac{1}{2}-b}
  \frac{\alpha^{-b}}
       {1 + \mu \alpha^{1-b}}
 -\frac{_1}{^{2b}}\,
  \min(u,v)^{\frac{1}{2}-b}\, \max(u,v)^{\frac{1}{2}+b}
 + o(\alpha^0)
\\
&=
  C_b\,
  (uv)^{\frac{1}{2}-b}\hspace{-0.2cm}
  \sum_{0\leq n\leq \frac{b}{1-b}}\hspace{-0.2cm}
   (-\mu)^n\, \alpha^{n(1-b)-b}
  -\frac{_1}{^{2b}}\,
  \min(u,v)^{\frac{1}{2}-b}\, \max(u,v)^{\frac{1}{2}+b}
 + o(\alpha^0)\,.
\end{align*}
This permits to infer the expansions
\begin{equation}
\label{eq:ltd}
e^{-tN_\mu}(u,v)\ =\ \
C_b\,
  (uv)^{\frac{1}{2}-b}
  \sum_{0\leq n<\frac{b}{1-b}}\hspace{-0.2cm}   
   \frac{(-\mu)^n}{\Gamma(b-n(1-b))}
    \,t^{b-n(1-b)-1}
   \,+\,o(t^{-1+\epsilon})
\end{equation}
for any $\epsilon>0$, and
\begin{multline}
\label{eq:ltf}
\ \hspace{0,5cm}L_\mu(t;u,v)
\ =\  
  C_b\,
  (uv)^{\frac{1}{2}-b}\hspace{-0.2cm}
  \sum_{0\leq n\leq \frac{b}{1-b}}
   \frac{(-\mu)^n}{\Gamma(1+b-n(1-b))}
    \,t^{b-n(1-b)}\hspace*{-1.4cm}
\\
  -\ \frac{_1}{^{2b}}\,
  \min(u,v)^{\frac{1}{2}-b}\, \max(u,v)^{\frac{1}{2}+b}
 \,+\, o(1)\,.\hspace{0.8cm}
\end{multline}
They give the large $t$ asymptotics relevant for the study
of the long time behavior of the tracer 2-point function
in the decaying and forced regime, respectively.

\subsection{Consequences for the tracer transport}
\label{ssect:tracer-transp}

Substituting the expansion \eqref{eq:ltd} into \eqref{eq:M-N} and then
into \eqref{eq:trans-proba}, one infers from \eqref{eq:2ptscr} that
the dominant terms in the free decay of the tracer 2-point function
are $r$-independent:
\begin{multline*}
\ \hspace{0.8cm}F(t,r)
\ =\ 
  (\frac{_{2-\xi}}{^2})^{2b-1} \tilde{D}_0^{b-1} C_b
  \sum_{0\leq n<\frac{b}{1-b}} 
    \frac{(-\mu)^n}{\Gamma(b-n(1-b))}
    \,t^{b-n(1-b)-1}
\\
\ \hspace{0.8cm}\cdot
  \,\Big(
    \int_0^\infty {r'}^{\,a-\xi}
     \,F(0,r')\,\dd r'\,+\,\tilde\lambda\,F(0,0)
  \Big)
  \ +\ o(t^{-1+\epsilon})
\,.\quad
\end{multline*}
Similarly, from \eqref{eq:ltf}, \eqref{eq:M-N}, \eqref{eq:trans-proba}
and, finally, \eqref{eq:2ptscfr}, we infer that in the forced
situation
\begin{multline}
\label{eq:t2pf}
  F(t,r)
=
\\
  (\frac{_{2-\xi}}{^2})^{2b-1} \tilde{D}_0^{b-1} C_b
  \sum_{0\leq n\leq \frac{b}{1-b}}
    \frac{(-\mu)^n}{\Gamma(1+b-n(1-b))}
    \,t^{b-n(1-b)}
\cdot
  \,\Big(
    \int\limits_0^\infty {r'}^{\,a-\xi}
    \,\chi(r')\,\dd r'\,+\,\tilde\lambda\,\chi(0)
  \Big)
\\
  -\,\frac{_1}{^{\tilde D_0(1-a)}}
  \Big(
    r^{1-a}\int\limits_0^r{r'}^{a-\xi}\,\chi(r') \,\dd r'
    +\int_r^\infty{r'}^{1-\xi}\chi(r') \,\dd r'
    \,+\, \tilde\lambda \, r^{1-a} \, \chi(0)
  \Big)
  \,+\,o(1)
\,.
\end{multline}
It is interesting to notice that the number of terms appearing in this
expansion is variable depending on the value of the integer part of
$\frac{b}{1-b}$.  In particular if $n(1-b)-b = 0$, i.e.\ $b = 1 -
\frac{1}{n+1}$ for some natural number $n$, then there is a
supplementary term of order 0 in the forced 2-point function.
\vskip 0.15cm

For $\tilde\lambda=0$, the result \eqref{eq:t2pf} reduces to the one
worked out in \cite{GV}.  The physics of the solutions with
$0<\tilde\lambda<\infty$ is not very different from the one for
$\tilde\lambda=0$ described in Sect.\ 3.2 and 3.3 of \cite{GV}.  The
tracer ``energy'' with density $\frac{1}{2}\theta^2$ is dissipated
but, in the forced case, also building up in the constant mode growing
like $t^b$.  Non-zero $\tilde\lambda$ brings subleading terms in this
buildup proportional to $t^{b-n(1-b)}$.  In the stationary state the
mean tracer energy density $\langle\frac{1}{2}\theta^2\rangle$ is
infinite.  The rate of pumping of the constant mode $\propto t^{b-1}$
goes, however, to zero.  The tracer energy dissipation rate equal to
$-\frac{1}{2}MF(t)|_{r=0}$ approaches at long times the stationary
value equal to the injection rate $\frac{1}{2}\chi(0)$ and the
stationary state exhibits a {\bf direct energy cascade}.  The tracer
2-point structure function approaches the stationary form
\begin{multline}
\label{eq:sticky-str-f}
\ \hspace{0.5cm}
  S_2(t,r)
\,=\,
  2(F(t,0)-F(t,r))
\\
\hspace{1cm}
\mathop{\longrightarrow}\limits_{t\to\infty}\ \ \ \ 
  \frac{_2}{^{\tilde D_0}}
  \Big[
    \int\limits_{0<r'<r''<r} \hspace{-0.4cm}
      {r''}^{\,-a_\xi} \, {r'}^{\,a_\xi-\xi} \,\chi(r')
    \,\dd r' \,\dd r''
    \,+\, \frac{_1}{^{1-a_\xi}} \, \tilde\lambda
       \, r^{1-a_\xi} \, \chi(0)\,
  \Big]
\,.\quad
\end{multline}
For small $r$ the integral on the right hand side scales as
$r^{2-\xi}$ while the term multiplying $\tilde{\lambda}$ is
proportional to $r^{1-a_\xi}$.  Since $1-a_\xi < 2-\xi$, a non-zero
$\tilde\lambda$ has an important effect. It changes the normal small
$r$ scaling $\propto r^{2-\xi}$ of the 2-point structure function
occurring for $\tilde{\lambda} = 0$ to the {\bf anomalous scaling}
proportional to the zero mode $r^{1-a_\xi}$ of $M$.
\vskip 0.15cm

Let us also briefly mention the case $\tilde\lambda = \infty$, i.e.\
the one with absorbing boundary condition and coalescence of
trajectories that was not covered above (the limits $t \to \infty$ and
$\tilde\lambda \to \infty$ do not commute!).
Using \eqref{expr:Bess-ccl}, the transition probabilities take in this
case the form
\begin{equation}
  P_\infty^t(r;\dd r')
\ =\ 
  e^{tM_\infty}(r,r') \,\dd r' \,+\,
  \Big[
    1 - \gamma(b, \frac{_{r^{2-\xi}}}{^{(2-\xi)^2\tilde D_0\,t}})
        \,\Gamma(b)^{-1}
  \Big]
  \,\delta(r') \,\dd r'
\,,
\nonumber
\end{equation}
the same as in the strongly compressible phase, see Eq.~(2.26) of
\cite{GV}.  The coefficient of $\delta(r')$ is obtained simply by
calculating the missing mass of $e^{tM_\infty}(r,r')$. 
For large time $t$, the 2-point function of the forced tracer becomes: 
\begin{multline*}
  F(t,r)
\ =\ 
  \Big(
    t \,-\,
    \frac{_{1}}{^{(2-\xi)^{2b}(1-b)\,\Gamma(1+b)}}
    {\tilde D}_0^{-b}\,t^{1-b}\,r^{1-a} \,+\,
    \frac{_{1}}{^{(2-\xi)^2(1-b)}}{\tilde D}_0^{-1}\,r^{2-\xi}
  \Big)
  \,\chi(0)
\\
\,+\,
  \frac{_1}{^{(1-a)\tilde D_0}}
  \Big(
        \int_0^r{r'}^{\,1-\xi}\,\chi(r')\,\dd r'\,+\,
        r^{1-a} \int_r^\infty{r'}^{\,a-\xi} \, \chi(r') \,\dd r'
  \Big)\,+\ o(1)\,.\hspace{0.7cm}
\end{multline*}
The behavior of this solution is similar to that of the strongly
compressible phase analyzed in Sect.\ 3.4 of \cite{GV} and quite
different from the one for finite $\tilde\lambda$.  The scalar energy
builds up linearly in time in the constant mode in an {\bf inverse
cascade} process and there is no persistent dissipation.  At
difference with the solution in the strongly compressible phase,
however, the tracer 2-point structure function does not reach a
stationary regime due to the contribution proportional to the zero
mode $r^{1-a}$ growing in time like $t^{1-b}$.

\section{Exact limit of the stationary 2-point tracer
structure function}
\label{sect:proof}

In the main part of the paper we have used in the calculations the
approximate forms of functions $f_1$ and $f_2$ instead of their exact
versions.  The question arises whether such calculations reproduce the
actual behavior of the limiting dispersion process, up to finite
renormalization of the glue parameter $\tilde\lambda$. Besides, one
would like to have exact expressions for some quantities, like the
mass at zero of the limiting speed measure or the limiting stationary
2-point structure function of the tracer. It turns out that it is
indeed possible to obtain precise formulae and that the asymptotic
behavior of the integrals that we calculated approximately differs
from the exact one by a (finite, non-zero) multiplicative
renormalization.
\vskip 0.15cm

In what follows, we shall need detailed information about functions
$f_{1,2}(r;l_\nu,l_\kappa)$ and their ratio $f_2(r;l_\nu,l_\kappa)/
f_1(r;l_\nu,l_\kappa)$ that is described in \appref{apx:f1-f2}.  In
particular we prove there the positivity for $r>0$ and the scaling
properties
\begin{equation}
\label{eq:scal12}
f_1(r;l_\nu)\,=\,l_\nu^{\xi}\,f_1(\frac{_{r}}{^{l_\nu}};1)\,,
\qquad
f_2(r;l_\nu)\,=\,l_\nu^{\xi-1}\,f_1(\frac{_{r}}{^{l_\nu}};1)
\end{equation}
of functions $f_{1,2}(r;l_\nu)\equiv f_{1,2}(r;l_\nu,0)$.  We also
establish two decompositions. The first one:
\begin{gather}
\label{ga:gad1}
  f_1(\rho;1)
=
  \tilde C_0\, \rho^2 f_1^\clubsuit(\rho)
=
  \tilde C_0\, \rho^2 + \rho^4 f_1^\spadesuit(\rho)
\\
\label{ga:gad2}
  f_2(\rho;1)
=
  a_2\, \tilde C_0\, \rho f_2^\clubsuit(\rho)
=
  a_2\, \tilde C_0\, \rho + \rho^3 f_2^\spadesuit(\rho)
\end{gather}
with $\,\lim\limits_{\rho \rightarrow 0}\, f_{1,2}^\clubsuit(\rho)=1$
will be used for small $\rho$.  The second one:
\begin{gather}
\label{ga:gad3}
  f_1(\rho;1)
=
  \rho^\xi f_1^\diamondsuit(\rho)
=
  \rho^\xi + \rho^{\xi-2} f_1^\heartsuit(\rho)
\\
\label{ga:gad4}
  f_2(\rho;1)
=
  a_\xi\,\rho^{\xi-1} f_2^\diamondsuit(\rho)
=
  a_\xi\,\rho^{\xi-1} + \rho^{\xi-3} f_2^\heartsuit(\rho)
\end{gather}
with $\,\lim\limits_{\rho \rightarrow \infty} \,
f_i^\diamondsuit(\rho) = 1$ will be employed for large $\rho$.  The
coefficients $a_2$ and $a_\xi$ are given by \eqref{def:a_xi}.  The
above decompositions are used in \appref{apx:f1-f2} to establish
bounds \eqref{bd:f_2/f_1;1} to \eqref{bd:f_2/f_1;3} on the ratio
$f_2(r;l_\nu,l_\kappa) / f_1(r;l_\nu,l_\kappa)$.  Given such bounds,
it is immediate to see that there exists a constant $B'$ such that,
for any $r>0$ and any $0 < l_\kappa < l_\nu<1$, the density $s(r)$ of
the natural scale measure given by \eqref{eq:scale-meas} is comprised
between $1/B'$ times and $B'$ times the approximate expressions listed
in \tabref{tab:s,m}.  Using additionally estimates \eqref{bd:f_1;1} to
\eqref{bd:f_1;3} from \appref{apx:f1-f2}, one infers that the same
statement holds for the density $m(r) = [f_1(r) s(r)]^{-1}$ of the
speed measure.  It is now easy to show the convergence of $s(r)$ and
$m(r)$ to $s_0(r)$ and $m_0(r)$ given by \eqref{def:s_0} and
\eqref{expr:m-dens-stoch} for $r>0$.  Controlling what happens around
$r=0$ will be the main difficulty.
\vskip 0.15cm

After the above preparation, we pass to the main topic of this
section: the proof of convergence, when $l_\nu,l_\kappa \to 0$, of the
stationary 2-point structure function of the forced passively advected
tracer.  We shall show that, under the condition \eqref{def:lambda}
with $\lambda<\infty$, the limit exists and corresponds to the
structure function computed directly at $l_\nu,l_\kappa=0$ with the
sticky boundary condition at zero, see \eqref{eq:sticky-str-f}.
Besides, we shall establish a precise relation between $\lambda$ and
the glue parameter $\tilde\lambda$ of the boundary condition
\eqref{expr:boundary-cond}.
\vskip 0.15cm

For $l_\nu,l_\kappa$ positive, the stationary 2-point structure
function is the unique solution vanishing at zero together with its
first derivative of the equation $M_{\nu,\kappa}S_2=2\chi$.  It is
given by the relation:
\begin{equation}
  \,S_2(r)
\,=\,
  \frac{_2}{^{\tilde D_0}} \int_0^r  s(r'')\,\dd r''
    \int_0^{r''}
      \hspace{-0.16cm}m(r')\, \chi(r')\,  \dd r'
\,=\,
  \frac{_2}{^{\tilde D_0}}
   \hspace{-0.35cm}\underset{0 < r' < r'' < r}{\int}
    \hspace{-0.45cm}s(r'')\, m(r')\, \chi(r')
  \,\dd r' \,\dd r''\,.
\nonumber
\end{equation}
The last integral may be cut up into six pieces according to the
positions of $r',r''$ with respect to $l_\kappa,l_\nu$.  The bounds
given for $s(r)$ and $m(r)$ indicate that in each domain the dominant
behavior of the contribution to $S_2(r)$ is estimated correctly by
using the rough forms in \tabref{tab:s,m}, up to multiplication by a
constant bounded independently of $l_\kappa, l_\nu$ and $r$.  We get
the following behaviors.  The domain $l_\nu < r' < r''$ gives a
contribution behaving at lowest order like $r^{2-\xi}$ for small $r$
and like $r^{1-a_\xi}$ for large $r$.  Domains $l_\kappa < r' < l_\nu
< r''$ and $r' < l_\kappa < l_\nu < r''$ both give terms of order
$l_\nu^{2-\xi+a_\xi-a_2} l_\kappa^{a_2-1} r^{1-a_\xi}$.  Finally the
contribution from domains $r'' < r' < l_\nu$ is always subdominant
compared to the former ones.  This suggests that the limiting value of
$S_2(r; l_\nu, l_\kappa)$ will depend on the limit of
$\,l_\nu^{2-\xi+a_\xi-a_2} l_\kappa^{a_2-1}\,$ as $l_\nu, l_\kappa$ go
to zero.  The only task left is to calculate the limit of $S_2(r;
l_\nu, l_\kappa)$ for $l_\nu^{2-\xi+a_\xi-a_2}
l_\kappa^{a_2-1}\to\lambda$ with $\lambda<\infty$.
\vskip 0.15cm

We already know that the domains with $0< r'' < r' < l_\nu$ will give
vanishing contributions (supposing that the others give a finite one).
Let us turn to the other three ones.  The easiest to handle is $l_\nu
< r'' < r' < r$.  Because of the bounds on $s(r)$ and $m(r)$ we can
immediately use the Dominated Convergence Theorem to obtain
\begin{multline*}
  \lim_{l_\nu,l_\kappa \rightarrow 0}
    \underset{l_\nu < r' < r'' < r}{\int}\hspace{-0.3cm}
      s(r'';l_\nu,l_\kappa)\, m(r';l_\nu,l_\kappa)\, \chi(r')
    \,\dd r' \,\dd r''
=\\
  \underset{0 < r' < r'' < r}{\int}\hspace{-0.3cm}
    s(r'';0,0)\, m(r';0,0)\, \chi(r')
  \,\dd r' \,\dd r''\ 
=
   \hspace{-0.1cm} \underset{0 < r' < r'' < r}{\int} \hspace{-0.3cm}
    {r''}^{\,-a_\xi}\, {r'}^{\,a_\xi - \xi}\, \chi(r')
  \,\dd r' \,\dd r''\,.\quad
\end{multline*}
Finally we turn to the two remaining domains covering the region ${0 <
r' < l_\nu < r'' < r}$.  The corresponding integral factorizes:
\begin{multline*}
    \underset{0 < r' < l_\nu < r'' < r}{\int}
      s(r'';l_\nu,l_\kappa)\, m(r';l_\nu,l_\kappa)\, \chi(r')
    \,\dd r' \,\dd r''
\\=\ 
    \left[
      {\int_{l_\nu}^r}
        s(r'';l_\nu,l_\kappa)
      \,\dd r''
    \right]
    \left[
      {\int_0^{l_\nu}}
        m(r';l_\nu,l_\kappa)\, \chi(r')
      \,\dd r' 
    \right].\quad
\end{multline*}
To the first factor, one may again apply the Dominated Convergence
Theorem to obtain
\begin{multline*}
  \lim_{l_\nu,l_\kappa \rightarrow 0}
    {\int_{l_\nu}^r}
      s(r'';l_\nu,l_\kappa)
    \,\dd r''
\\=\ 
  {\int_0^r}
    s(r'';0,0)
  \,\dd r''
\ =\ 
  \int_0^r \left(\frac{r_0}{r''}\right)^{a_\xi}  \,\dd r''
\ =\ 
  \frac{1}{1-a_\xi}\,r_0^{a_\xi}\,r^{1-a_\xi}.\quad
\end{multline*}
Control of the second factor, that converges to the mass at zero of
the limiting speed measure multiplied by $\chi(0)$, is the crucial
element of the argument and it requires technical work.  We postpone
it to \appref{apx:m(0)}. We prove there that for $\lambda<\infty$,
\begin{equation}
\label{eq:m(0)}
  \lim\limits_{l_\nu,l_\kappa\mathop{\to}\limits^\lambda 0}
    \ \int_0^{l_\nu}
    m(r';l_\nu,l_\kappa)\, \chi(r')
  \,\dd r'
\ =\   
   r_0^{-a_\xi}\,Y\,\lambda\,\chi(0)
\quad 
\end{equation}
with 
\begin{multline}
\label{eq:Y}
\ \quad Y\ =\ \frac{_1}{^4}
  \left(\frac{_1}{^2}\tilde C_0\right)^{-\frac{a_2+1}{2}}
  B(\frac{_d}{^2},\frac{_{1-a_2}}{^{2}})\,
\\
\cdot\,
\exp\left[
    -
    \int_{0}^{1}
      \frac{
        \rho
        \left[
          f_2^\spadesuit(\rho)
          - a_2 f_1^\spadesuit(\rho)
        \right]
      }{
        \tilde C_0
        f_1^\clubsuit(\rho)
      }
    \,\dd \rho
    -
    \int_1^\infty
      \frac{
        f_2^\heartsuit(\rho)
        - a_\xi f_1^\heartsuit(\rho)
      }{
        \rho^3
        f_1^\diamondsuit(\rho)
      }
    \,\dd \rho
  \right]\quad
\end{multline}
where $\,l_\nu,l_\kappa\mathop{\to}\limits^\lambda 0\,\,$ denotes the 
limit $l_\nu,l_\kappa \to 0$ under condition \eqref{def:lambda} and $B(x,y)$ 
is the Euler Beta function. The constant $\,Y\,$ is positive.  
The mass at zero of the limiting
speed measure is then equal to $r_0^{-a_\xi} Y \,
\lambda$ which should be contrasted with the approximate formula
\eqref{expr:m-mass-stoch}.  Given the above result, the contribution
of the region $\,0<r'<l_\nu<r''<r\,$ to $\,S_2(r)\,$ becomes
$$
  \frac{2\,Y}{\tilde D_0(1-a_\xi)}\,\lambda\, r^{1-a_\xi} \,\chi(0)\,.
$$
The final outcome is the relation:
\begin{equation}
\lim_{l_\nu,l_\kappa\mathop{\to}\limits^\lambda 0}\ S_2(r)\ = 
\ \frac{_2}{^{{\tilde D}_0}}\Big(
\underset{0 < r' < r'' < r}{\int}
    \hspace{-0.2cm}{r''}^{\,-a_\xi} {r'}^{\,a_\xi - \xi}\, \chi(r')
  \,\dd r' \,\dd r''
  \ +\ \frac{_{Y}}{^{1-a_\xi}} \,\lambda\, 
       r^{1-a_\xi}\,\chi(0)\,\Big)\quad  
\nonumber
\end{equation}
which agrees with expression \eqref{eq:sticky-str-f} of
\secref{ssect:tracer-transp} if we put $\tilde\lambda = Y \lambda$.
This means that the structure function obtained in the
$\,l_\nu,l_\kappa\mathop{\to}\limits^\lambda0\,$ limit coincides with
the one that is obtained directly using the sticky process with the
generator $M_{\tilde\lambda}$ for the value
\begin{equation}
\label{expr:glue-param-3}
  \tilde\lambda
=
  Y \lambda
\end{equation}
of the glue parameter.  
Note that the effect of using the exact versions of $f_1,f_2$ instead
of the approximate versions of table \tabref{tab:f1-f2} manifests
itself only in the change of the proportionality constant between 
$\lambda$ and the glue parameter $\tilde\lambda$, i.e.\ in a finite
multiplicative renormalization of $\tilde\lambda$.
\vskip 0.15cm

In view of all that has been said, it is natural to expect that the
entire Lagrangian pair dispersion process behaves in the
$\,l_\nu,l_\kappa\mathop{\to}\limits^\lambda0\,$ limit as the
diffusion process with the generator $M_{\tilde\lambda}$ for the glue
parameter $\tilde\lambda$ given by \eqref{expr:glue-param-3}.  In
particular, such a process is equivalent in the coordinate
\eqref{def:Bessel-coord} to the Bessel process of parameter
$-b_{\xi,\wp}$ with a sticky behavior at zero.

\section{Conclusions}

We have analyzed in this article the small viscosity, small
diffusivity behavior of the Lagrangian dispersion in the Kraichnan
model with intermediate compressibility degree
$\frac{d-2}{2\xi}+\frac{1}{2}<\wp<\frac{d}{\xi^2}$.  In this interval,
the Lagrangian trajectories may separate fast due to the spatial
roughness of velocities but may also come close due to the trapping
effects of compressibility.  As first suggested in \cite{EVDE1,EVDE2},
we have discovered different possible asymptotic regimes of the
Lagrangian flow, depending on the limiting behavior of the Prandtl
number when the viscous and diffusive cut-off scales $l_\nu$ and
$l_\kappa$ are taken to zero.  This arbitrariness reflects the
frustration of the particles unable to choose between opposites trends
of life.  Specifically, we have argued that there are different limits
of the dispersion process depending on the behavior of the combination
$$
  l_\nu^{a_\xi+2-\xi-a_2}\,  l_\kappa^{a_2-1}
\ =\ 
  {\rm const}.\ l_\nu^{a_\xi+1-\xi} (\mathit{Pr})^{\frac{1-a_\xi}{2}}
$$
of the cutoff scales.  If this combination goes to zero when $l_\nu$
and $l_\kappa$ are sent to zero, the resulting dispersion process is
that of trajectories instantaneously reflecting off each other upon
hitting.  If it goes to infinity, the trajectories coalesce when they
meet, behaving similarly as in the strongly compressible regime.
Finally, if the above combination goes to a finite limit $\,\lambda\,$
when $l_\nu,l_\kappa\to0$ (which sends to infinity the Prandtl number
at a specific pace) then the resulting dispersion process exhibits the
sticky or slowly reflecting behavior with $\lambda$ proportional to
the amount of ``glue'' keeping the particles together.  Such a
behavior leaves a visible imprint on the passive advection of tracers
in the subleading contributions to the tracer energy condensation in
the zero wave number and, even more dramatically, by generating
anomalous scaling of the stationary 2-point structure function of the
tracer.  The analysis in the main part of the paper was based on
approximate calculations of the asymptotic behavior of the natural
scale and speed measure of the Lagrangian dispersion process and of
the eigenfunctions of its generator.  To set the results on a firmer
ground we have also shown rigorously that the finite $l_\nu,l_\kappa$
stationary 2-point structure function of the tracer converges, under
the limit $l_\nu,l_\kappa \to 0$ with finite $\lambda$, to the
stationary structure function obtained directly from the sticky
dispersion process.  This argument allowed to fix exactly the
proportionality constant between $\lambda$ and the value of the glue
parameter.  Undoubtedly, with a little more work controlling the
convergence of resolvents of $\,M_{\nu,\kappa}\,$ to that of
$M_{\tilde\lambda}$ one should be able to prove, along the lines of
\cite{FW} that the laws of the dispersion processes for
positive $l_\nu,l_\kappa$ converge to the law of the sticky process.
\vskip 0.15cm

The main open problem, untouched by our analysis, is the construction
of $N$-particle processes corresponding to the sticky behavior of the
two-particle dispersion.  In particular it would be interesting to
know whether the amount of two-particle glue is the only parameter
that labels possible Lagrangian flows in the moderately compressible
phase of the Kraichnan model.  The Dirichlet form approach used in
\cite{LeJR3} in the 1-dimensional $\xi=0$ case to tackle such
questions is unavailable in the other instances, at least in its
classical form, due to the lack of symmetry of the generators of the
$N$-particle processes.  Further open questions of fundamental
importance concern possible occurrence of sticky Lagrangian flows in
more realistic velocity ensembles exhibiting fully developed
turbulence.
 
\vskip 1cm

\noindent{\bf\Large Appendices}
\vskip -0.6truecm
\ 

\appendix

\section{Briefly on one-dimensional diffusion processes}
\label{apx:1d-dif}

We collect here some facts, used in the main text, about
one-dimensional diffusion processes.  The reader may wish to consult
the relevant literature (e.g.\ \cite{Breiman,RW}) for an extensive
treatment.
\vskip 0.15cm

Let $X(t)$ be a stochastic diffusion process on $\RR_+$.  The
Kolmogorov backwards evolution operator $P^t$ of the process, acting
on bounded continuous functions $f$ defined on $\RR_+$, is given by
$$
  (P^t f)(r)
\,=\,
  \EE_r f(X(t))
\,=\,
  \int f(r') P^t(r, \dd r')
$$
where we define $\EE_r f(X(t)) \equiv \EE(f(X(t)) \mid X(0) = r)$.  The
measures $P^t(r,\dd r')$ giving the kernels of operators $P^t$ are the
transition probabilities of the process.  The family $(P^t)_{t \in
\RR_+}$ forms a one-parameter semigroup whose generator $\partial_t
P^t |_{t=0}$ we shall denote by $M$. In general $M$ is a second order
differential operator.  The transition probabilities verify the PDE
$$
  \partial_t P^t(r, \dd r')
=
  M(r) P^t(r, \dd r')\,.
$$
\vskip 0.1cm

Any regular diffusion process may be derived from Brownian motion
by an adequate change of variables defined in terms of the natural
scale and the speed measure associated to the process.  The natural scale
is defined as the unique, up to affine transformations, continuous
strictly increasing function $S$ such that $S(X(t))$, i.e. $X(t)$
considered in the new spatial coordinate $S$, is a martingale.  $S$ is
called the natural scale and $s(r)= \frac{\dd S(r)}{\dd r}$ the
density of the natural scale measure with respect to the coordinate
$r$ on $\RR_+$.
\vskip 0.15cm

If the range of $S$ is the whole real line then the process $S(X(t))$,
for $X(0)$ fixed, has the same distribution as an appropriately
time-changed Brownian motion starting from $S(X(0))$.  This means that
if $W$ is a Brownian motion starting from $S(X(0))$, then $S(X(t))$
has the same law as $W(\tau(t))$ for some function $\tau(t) =
\tau(t;W)$ which depends also on the realization of the Brownian path.
Let us denote by $L(\tau,S;W)$ the local time of the Brownian path $W$
at the point $S$ up to instant $\tau$. Formally, $L(\tau,S;W) =
\int_0^\tau \delta(S-W(\sigma)) \, \dd\sigma$.  The speed measure of
the process $X(t)$ is defined as the unique positive measure $\dd
m(S)$ on $\RR$
such that the relation between $t$ and $\tau$ may be written as
$$
  \tau(t;W)
= 
  \inf\left\{
    \sigma : \int L(\sigma,S;W) \,  \dd m(S) > t
  \right\}
$$
or, equivalently,
$$
  t(\tau;W)
=
  \int L(\tau,S;W) \, \dd m(S)\,.
$$
Intuitively, the speed measure expresses how much time the
(time-changed) Brownian motion needs to advance ``one step'' at a
given point in space.  The larger the (density of the) speed measure
at some point, the slower the (time-changed) Brownian motion advances.
\vskip 0.15cm

The generator $M$ of the Kolmogorov backward evolution semigroup may
be written in terms of the densities $m(S) = \frac{\dd m(S)}{\dd S}$
or $\,m(r) = \frac{\dd m(r)}{\dd r}$ and $s(r)$ as
$$
  M
\,=\,
 \frac{1}{2m(S)} \partial_S^2
\,=\,
  \frac{1}{2m(r)} \partial_r 
  \circ
  \frac{1}{s(r)} \partial_r 
\,.
$$
\vskip 0.1cm

Similarly, if the range of $S$ is a positive half-line, which may be
chosen as $\RR_+$, then analogous statements hold with the Brownian
motion $W$ replaced by the Brownian motion reflecting at zero, i.e.
by $|W(t)|$. In any case, the behavior of the process at the boundary
point $r=0$ may be classified according to Feller's criteria
expressible in terms of the natural scale and the speed measure.  When
$r=0$ is a regular boundary point, the behavior of the process at this
point depends on the its mass $m(\{0\})$ w.r.t the speed measure
and is reflected in the boundary condition for the generator $M$.
\vskip 0.15cm

Finally let us recall from \cite{IMcK} the following formula for
hitting probabilities of some process $X_t$.  Denote by $H_{r'}$ the
hitting time of the process at some point $r'$.  Then for $\alpha >
0$
\begin{equation}
\label{expr:hitting}
  \EE_r(e^{-\alpha H_{r'}})
=
  \frac{\phi(r)}{\phi(r')}
\end{equation}
where $\phi$ is the solution of $(M-\alpha) \phi = 0$ that verifies
the correct boundary condition either on the left (i.e.\ at 0 if the
process is defined on the half-line) if $r < r'$ or on the right
(i.e.\ at infinity if the process is defined on the half-line) if $r >
r'$.

\section{A pathological case of convergence}
\label{apx:patho}

Let $L$ be an arbitrary length and $l$ a small scale that we shall
send to zero.  A diffusion process may be specified by giving its
natural scale and its speed measure.  Consider such a process on
$\RR_+$ whose natural scale is
$$
  S(r)
=
  \begin{cases}
    l-L + \frac{L}{l} r & \quad{\rm for}\quad r \in [0,l]\,, \\
    r                   & \quad{\rm for}\quad r \in [l,\infty)
  \end{cases}
$$
so that the density of the natural scale measure w.r.t.\ $r$ is
$$
  s(r)
=
  \begin{cases}
    \frac{L}{l} & \quad{\rm for}\quad r \in [0,l]\,, \\
    1                   & \quad{\rm for}\quad r \in [l,\infty)\,.
  \end{cases}
$$
Let us take for (the density w.r.t.\ $r$ of) the speed measure
$$
  m(r)
=
  \begin{cases}
\frac{L}{l} & \quad{\rm for}\quad r \in [0,l]\,, \\
     1       & \quad{\rm for}\quad r \in [l,\infty)\,.
  \end{cases}
$$
We see that taking the limit $l \to 0$ for $r>0$ we get $s_0(r) = 1$
corresponding to $S_0(r)= r$, however $\lim\limits_{l \to 0} S(0) =
-L$.  This is not very different from the case studied in the
\secref{sect:stoch-proc}, where $S(0) = -\infty$ as long as there is
finite regularization.  For the limit of the speed measure we get
$m_0(r) = L\delta(r) + 1$.
\vskip 0.15cm

Let us now pass to the natural scale first.  Then $m(S) =
\frac{m(r)}{s(r)} = 1$ for all $S$.  Thus $m_0(S) = 1$.  This is of
course incompatible with the previous result.
\vskip 0.15cm

The explanation of this phenomenon is the following.  To the right of
$l$ the process is just Brownian motion.  To the left of $l$ it is
Brownian motion on a segment of length $L$ (with reflecting left end)
``squeezed'' into the segment $[0,l]$.  Thus when we take $l$ to zero,
the limiting process will be a Brownian motion on $\RR_+$ which
``sees'' an additional segment of length $L$ to its left.  This is of
course a non-Markovian boundary condition.

\section{Asymptotic behavior of eigenfunctions}
\label{apx:exp}

Starting from formula \eqref{eq:chain2} we calculate the limit of
$c_E^+/c_E^-$ when $l_\kappa,l_\nu \rightarrow 0$.  We assume that
$l_\kappa < l_\nu$, but not necessarily $l_\kappa \ll l_\nu$.  The
calculation can be done in the usual way by expanding $c_E^+/c_E^-$ in
a multivariate power series.  First we expand every term in
\eqref{eq:chain2} to the {\bf minimal orders} in $l_\nu,l_\kappa$.
That is, we expand the expressions into a power series (and possibly
powers of logarithms) in $l_\nu,l_\kappa$ and keep all the terms such
that there is no term of smaller order simultaneously in $l_\nu$ and
in $l_\kappa$.
\vskip 0.15cm

We list below the expansion to minimal orders in $l_\kappa,l_\nu$ of
all terms appearing in \eqref{eq:chain2}:
\begin{alignat*}{3}
& g_1^- (l_\kappa)
\sim
  1
\quad&\quad
& {g_1^-}' (l_\kappa)
\sim
  -\frac{E}{d} l_\nu^{2-\xi} l_\kappa^{-1} 
\\
& \gamma^-
\sim
  \frac{E}{1-a_2} l_\nu^{2-\xi}
\quad&\quad
& \gamma^+
\sim
  1 - a_2
\\
& g_2^- (l_\kappa)
=
  (l_\kappa)^{\gamma^-}
\quad&\quad
& {g_2^-}' (l_\kappa)
=
  \gamma^- (l_\kappa)^{\gamma^- - 1}
\sim
  \frac{E}{1-a_2} l_\nu^{2-\xi} l_\kappa^{-1} (l_\kappa)^{\gamma^-}
\\
& g_2^+ (l_\kappa)
=
  (l_\kappa)^{\gamma^+}
\sim
  l_\kappa^{1-a_2}
\quad&\quad
& {g_2^+}' (l_\kappa)
=
  \gamma^+ (l_\kappa)^{\gamma^+ - 1}
\sim
  (1-a_2) l_\kappa^{-a_2}
\\
& g_2^- (l_\nu)
=
  (l_\nu)^{\gamma^-}
\sim
  e^{\frac{E}{1-a_2} l_\nu^{2-\xi} \ln l_\nu}
\sim
  1
\quad&\quad
& {g_2^-}' (l_\nu)
=
  \gamma^- (l_\nu)^{\gamma^- - 1}
\sim
  \frac{E}{1-a_2} l_\nu^{1-\xi}
\\
& g_2^+ (l_\nu)
=
  (l_\nu)^{\gamma^+}
\sim
  l_\nu^{1-a_2}
\quad&\quad
& {g_2^+}' (l_\nu)
=
  \gamma^+ (l_\nu)^{\gamma^+ - 1}
\sim
  (1-a_2) l_\nu^{-a_2}
\\
& g_3^- (l_\nu)
\sim
  1
\quad&\quad
& {g_3^-}' (l_\nu)
\sim
  - \frac{(2-\xi)^2 E}
         {4 (1 - \xi + a_\xi)}
  l_\nu^{1-\xi}
\\
& g_3^+ (l_\nu)
\sim
  l_\nu^{1-a_\xi}
\quad&\quad
& {g_3^+}' (l_\nu)
\sim
  (1-a_\xi) l_\nu^{-a_\xi}
\end{alignat*}
Note that $(l_\kappa)^{\gamma^-}$ cannot be further expanded without
any additional hypothesis on the relative behaviors of $l_\nu$ and
$l_\kappa$ because both the base and the exponent go to zero and one
depends on $l_\kappa$ the other on $l_\nu$.  All we can say is that
$l_\kappa^\epsilon = o((l_\kappa)^{\gamma^-})$ for any $\epsilon > 0$,
and $(l_\kappa)^{\gamma^-} < 1$ when $l_\kappa < 1$.  We shall keep
$(l_\kappa)^{\gamma^-}$ as it is in the expansions.
\vskip 0.15cm

Calculations to minimal order with expansions into multivariate power
series are a little trickier then with univariate expansions.  Even if
the coefficients of some but not all monomials in the expansion
simplify to zero at intermediate stages then possibly higher order
terms should be taken into account because they could give rise to
terms of minimal order\footnote{%
        Take for example the sum of two polynomials $(X + XY + Y^2) +
        (-X + Y^2)$.  The sum of the minimal order expansions is $(X
        + Y^2) + (-X + Y^2) = 2 Y^2$ but the minimal order expansion
        of the sum is of course $XY + 2 Y^2$.}.
This precaution is implicit in the computations.  However, as long as
there is only one term of minimal order (and it does not simplify
out), no special care is needed.
\vskip 0.15cm

After having determined above the behavior of each term, we may now
calculate the sub-products of \eqref{eq:chain2} to minimal order.
First evaluate
\begin{multline*}
  \ \qquad\begin{pmatrix}
     {g_3^-}' (l_\nu) &  -g_3^-   (l_\nu) \\
    -{g_3^+}' (l_\nu) &   g_3^+   (l_\nu)
  \end{pmatrix}
  \begin{pmatrix}
     g_2^+   (l_\nu) &  g_2^-   (l_\nu) \\
    {g_2^+}' (l_\nu) & {g_2^-}' (l_\nu)
  \end{pmatrix}
\\
\sim\,  \begin{pmatrix}
    -(1-a_2) \, l_\nu^{-a_2} &
    -\left[
       \frac{(2-\xi)^2}{4(1 - \xi + a_\xi)} +
       \frac{1}{1-a_2}
     \right] E \,
    l_\nu^{1-\xi}
  \\
    (a_\xi - a_2) \, l_\nu^{1 - a_2 - a_\xi} &
    -(1-a_\xi) \, l_\nu^{-a_\xi}
  \end{pmatrix}.\qquad
\end{multline*}
It is straightforward to check that every coefficient is different
from zero, except for $a_\xi - a_2$ which may be equal to zero.  So there 
the expansion has to be pushed further, but since each following term
is of higher order both in $l_\nu$ and in $l_\kappa$,
we may simply replace $(a_\xi - a_2) \, l_\nu^{1 - a_2 - a_\xi}$ by
$O(l_\nu^{1 - a_2 - a_\xi})$.
\,Next evaluate
\begin{equation*}
  \begin{pmatrix}
     {g_2^-}' (l_\kappa) &  -g_2^-   (l_\kappa) \\
    -{g_2^+}' (l_\kappa) &   g_2^+   (l_\kappa)
  \end{pmatrix}
  \begin{pmatrix}
     g_1^-   (l_\kappa) \\
    {g_1^-}' (l_\kappa)
  \end{pmatrix}
\sim
  \begin{pmatrix}
    \left[
      \frac{1}{1-a_2} +
      \frac{1}{d}
    \right] E \,
    l_\nu^{2-\xi} l_\kappa^{-1} (l_\kappa)^{\gamma^-} \\
    -(1-a_2) \, l_\kappa^{-a_2}
  \end{pmatrix}.
\end{equation*}
Again, it is straightforward to check that every coefficient is
different from zero.
\,We may multiply together (exactly) the preceding two
subexpressions to arrive at
\begin{equation}
\label{eq:quot-c_E}
  \begin{pmatrix}
    c_E^+ \\
    c_E^-
  \end{pmatrix}
\,\mathop{\propto}\limits_{^\sim}\,
  \begin{pmatrix}
      -\frac{d+1-a_2}{d} E \,
      l_\nu^{2 - \xi - a_2} l_\kappa^{-1} (l_\kappa)^{\gamma^-}
    +
      \left[
        1 +
        \frac{(2-\xi)^2 (1-a_2)}
             {4 (1 - \xi + a_\xi)}
      \right] E \,
      l_\nu^{1-\xi} l_\kappa^{-a_2}
  \\
      O\left(
        l_\nu^{3 - a_2 - a_\xi - \xi}
        l_\kappa^{-1} (l_\kappa)^{\gamma^-}
      \right)
    +
      (1-a_2) (1-a_\xi) \,
      l_\nu^{-a_\xi}
      l_\kappa^{-a_2}
  \end{pmatrix}.
\end{equation}
Now all the terms are of minimal order.  Additionally, we have
\begin{gather}
\label{eq:cmp1}
  l_\nu^{3 - a_2 - a_\xi - \xi} l_\kappa^{-1} (l_\kappa)^{\gamma^-}
=
  l_\nu^{1-a_\xi}
  (l_\nu^{2 - \xi - a_2} l_\kappa^{-1} (l_\kappa)^{\gamma^-})
=
  o(l_\nu^{2 - \xi - a_2} l_\kappa^{-1} (l_\kappa)^{\gamma^-})\,,
\\
\label{eq:cmp2}
  l_\nu^{1-\xi} l_\kappa^{-a_2}
=
  l_\nu^{a_\xi+1-\xi} (l_\nu^{-a_\xi} l_\kappa^{-a_2})
=
  o(l_\nu^{-a_\xi} l_\kappa^{-a_2})\,.\hspace{0.42cm}
\end{gather}
\vskip 0.1cm

Let us first suppose that $l_\nu^{2 - \xi - a_2} l_\kappa^{-1}
(l_\kappa)^{\gamma^-} = o(l_\nu^{-a_\xi} l_\kappa^{-a_2})$.  In this
case $c_E^+/c_E^-$ goes to zero.  This is so even if in
\eqref{eq:quot-c_E} the expansion for $c_E^+$ cancels out, meaning
that the expansion should be pushed further.  Indeed, subsequent terms
in the expansion are of higher order thus asymptotically smaller.  On
the other hand, there can be no cancellation in the expansion of
$c_E^-$ because \eqref{eq:cmp1} and the hypothesis of the present
paragraph combine to $l_\nu^{3 - a_2 - a_\xi - \xi} l_\kappa^{-1}
(l_\kappa)^{\gamma^-} = o(l_\nu^{-a_\xi} l_\kappa^{-a_2})$.

Conversely, if we suppose $l_\nu^{-a_\xi} l_\kappa^{-a_2} = o(l_\nu^{2
- \xi - a_2} l_\kappa^{-1} (l_\kappa)^{\gamma^-})$, then $c_E^+/c_E^-$
goes to $\infty$.  This even if in \eqref{eq:quot-c_E} the expansion
for $c_E^-$ cancels out, meaning that the expansion should be pushed
further.  Once again, subsequent terms in the development are of
higher order thus asymptotically smaller.  And now it is in the
development of $c_E^+$ that there can be no cancellation because
\eqref{eq:cmp2} and the hypothesis of the present paragraph combine
to $l_\nu^{1-\xi} l_\kappa^{-a_2} = o(l_\nu^{2 - \xi - a_2}
l_\kappa^{-1} (l_\kappa)^{\gamma^-})$.
\vskip 0.15cm

In consequence, if $c_E^+/c_E^-$ should have a finite non-zero
limit, then $\,l_\nu^{2 - \xi - a_2} l_\kappa^{-1}
(l_\kappa)^{\gamma^-}$ and $l_\nu^{-a_\xi} l_\kappa^{-a_2}$ must be of
the same order.  In particular they are much bigger than the other two
terms so there are no cancellations in \eqref{eq:quot-c_E}.  Also it
is easy to see that in this case $l_\kappa$ cannot decrease faster
than some power of $\,l_\nu$.  If we write $l_\nu^{-a_\xi}
l_\kappa^{-a_2} = O(l_\nu^{2 - \xi - a_2} l_\kappa^{-1}
(l_\kappa)^{\gamma^-})$ then, using $(l_\kappa)^{\gamma^-} = O(1)$, we
have $l_\nu^{\frac{2-\xi-a_2+a_\xi}{1-a_2}} = O(l_\kappa)$.  This
implies $(l_\kappa)^{\gamma^-} \sim 1$.  We may then conclude that
$$
  \frac{c_E^+}{c_E^-}
\sim
   -\frac{d+1-a_2}{(1-a_2) (1-a_\xi) d}\, E \,
  l_\nu^{2-\xi-a_2+a_\xi} l_\kappa^{a_2-1}
\,.
$$
This expression has a finite limit if and only if
$\,l_\nu^{2-\xi-a_2+a_\xi} l_\kappa^{a_2-1}$ goes to some finite limit
$\lambda$, just as in \eqref{def:lambda}.  We obtain this way formula
\eqref{eq:quot-coeff}.  In fact our proof shows that
\eqref{eq:quot-coeff} is valid also if $\lambda$ is zero or infinite.

\section{Spectral analysis of $N_\mu$}
\label{apx:spec-meas-N}

The (operator valued) spectral measure for the operator $N_\mu$ may be
evaluated with the help of the formula
\begin{equation}
  E_\mu(\mathcal{B})
=
  \frac{1}{2\pi i}
  \lim_{\epsilon \rightarrow 0}
    \int_{\mathcal{B}}
       [ G_{\mu,\alpha+i\epsilon}
        -G_{\mu,\alpha-i\epsilon}]
    \,\dd\alpha
\nonumber
\end{equation}
where $G_{\mu,\alpha}$ is the resolvent
$$
  G_{\mu,\alpha}
\equiv
  (N_\mu - \alpha)^{-1}
$$
and $\mathcal{B}$ is a (Borel) subset of $\RR$.
The kernel $G_{\mu,\alpha}(u,v)$ of $G_{\mu,\alpha}$ is determined by 
demanding that for each $v \in \RR_+$, as a function of $u\in\RR_+$, it 
is in the domain of $N_\mu$ and that it satisfies the equation
\begin{equation}
\label{eq:resolvent}
  (N(u) - \alpha)\, G_{\mu,\alpha} (u,v)
=
  \delta(u-v)\,.
\end{equation}
To calculate $G_{\mu,\alpha}(u,v)$ we note that for $u \neq v$ we simply
have $\,(N(u) - \alpha) G_{\mu,\alpha} (u,v) = 0$.  Viewed
as a function of $u$, $G_{\mu,\alpha} (u,v)$ verifies at zero the
boundary condition \eqref{expr:N-bc0} and should go to zero at
infinity. A solution of $(N_\mu - \alpha) \phi(u) = 0$ satisfying the
correct boundary condition at zero is
\begin{equation}
\label{expr:ef-N-bc0}
  \phi_{\mu,\alpha}(u)
=
  u^{\frac{1}{2}}
  \left[
                          I_{-b}(\sqrt{-\alpha}\, u)
    + \mu (-\alpha)^{1-b} I_b   (\sqrt{-\alpha}\, u)
  \right],
\qquad
  \alpha \notin \RR_+
\,.
\end{equation}
Here $I_{\pm b}$ is the modified Bessel function of parameter $\pm b$.
The square-root is taken with its principal definition.  
Similarly, the solution of $\,(N - \alpha) \psi(u) = 0\,$ decaying at
infinity is
\begin{equation}
\label{expr:ef-N-bcinfty}
  \psi_{\alpha}(u)
=
  u^{1/2} K_b(\sqrt{-\alpha} u)\,,
\qquad
  \alpha \notin \RR_+
\end{equation}
where $K_b$ is the modified Bessel function of the second kind of
parameter $b$.  
With the correct matching at $u=v$ to assure that \eqref{eq:resolvent}
is satisfied, we obtain
\begin{equation}
\label{expr:G}
  G_{\mu,\alpha}(u,v)
=
  \begin{cases}
    \frac{\phi_{\mu,\alpha}(u) \psi_{\alpha}(v)}
         {w_{\mu,\alpha}}
    \qquad{\rm if}\quad u \leq v\,,
  \\[3mm]
    \frac{\psi_{\alpha}(u) \phi_{\mu,\alpha}(v)}
         {w_{\mu,\alpha}}
    \qquad{\rm if}\quad u \geq v
  \end{cases}
\end{equation}
where $w_{\mu,\alpha}$ is the Wronskian of $\phi_{\mu,\alpha}$ and
$\psi_{\alpha}$, i.e.
$$
  w_{\mu,\alpha}
\equiv
    \phi'_{\mu,\alpha}(z) \psi_{\alpha}(z)
  - \phi_{\mu,\alpha}(z) \psi'_{\alpha}(z)
$$
which is independent of $z$.
To evaluate the Wronskian, we may use the asymptotic
expansions  for $z \rightarrow +\infty$ 
\begin{equation}
  I_\nu(z)
=
  \frac{e^z}
       {\sqrt{2\pi z}}
  (1 + O(\frac{1}{z}))\,,
\qquad
  K_\nu(z)
=
  \sqrt{\frac{\pi}{2z}}
  e^{-z}
  (1 + O(\frac{1}{z}))\,,
\nonumber
\end{equation}
see Eq.~8.451.6,7 of \cite{GR}. The result is
$$
  w_{\mu,\alpha}
=
  1 + \mu (-\alpha)^{1-b}\,.
$$

Next we have to calculate the discontinuity in $G_{\mu,\alpha}(u,v)$
along the cut for $\alpha \in \RR_+$.  It will be convenient to write
$-(\alpha \pm i 0) = e^{\mp \pi i} \alpha$.  This will automatically
give the correct determination of every function.  For $u<v$
\begin{multline}
\ \quad  G_{\mu,\alpha \pm i 0} (u,v)
\ =\ 
\frac{1}{1+\mu \alpha^{1-b} e^{\mp \pi i (1-b)}}
\,
u^{1/2}
  \Big[
      I_{-b}(e^{\mp \frac{\pi i}{2}} \sqrt{\alpha} u)
\\
+\ \mu e^{\mp \pi i (1-b)} \alpha^{1-b}
      I_b   (e^{\mp \frac{\pi i}{2}} \sqrt{\alpha} u)
  \Big]
\,
 v^{1/2}
  K_b(e^{\mp \frac{\pi i}{2}} \sqrt{\alpha} v)\,.\quad
\nonumber
\end{multline}
Now we employ the general formula 
$$
  K_\nu(z)
=
  \frac{\pi}{2}
  \frac{I_{-\nu}(z) - I_\nu(z)}
       {\sin \nu\pi}
\qquad{\rm for}\quad
  \nu \notin \ZZ\,,
$$
see Eq.~8.485 of \cite{GR}, to get
\begin{multline}
\ G_{\mu,\alpha \pm i 0} (u,v)
\,=\, 
\frac{1}{1+\mu \alpha^{1-b} e^{\mp \pi i (1-b)}}
 u^{1/2}
  \Big[
      I_{-b}(e^{\mp \frac{\pi i}{2}} \sqrt{\alpha} u)
\\
+\,\mu e^{\mp \pi i (1-b)} \alpha^{1-b}
      I_b   (e^{\mp \frac{\pi i}{2}} \sqrt{\alpha} u)
  \Big]\, v^{1/2}
  \frac{\pi}{2 \sin b\pi}
  \left[
      I_{-b}(e^{\mp \frac{\pi i}{2}} \sqrt{\alpha} v)
    - I_b   (e^{\mp \frac{\pi i}{2}} \sqrt{\alpha} v)
  \right].\ \ 
\nonumber
\end{multline}
The relations
\begin{equation}
  I_\nu(e^{\frac{\pi i}{2}} z)
=
  e^{\frac{\pi i}{2} \nu}
  J_\nu(z)\,,\qquad
  I_\nu(e^{-\frac{\pi i}{2}} z)
=
  e^{-\frac{\pi i}{2} \nu}
  J_\nu(z)
\nonumber
\end{equation}
following from Eqs.,8.406.1 and 8.476.1 of \cite{GR} 
permit to move the determination from the function argument to a
multiplicative coefficient.  The final expression is
\begin{align*}
  G_{\mu,\alpha \pm i 0} (u,v)
\ =\ 
& \frac{\pi}{2 \sin b\pi}
  \frac{1}{1+\mu \alpha^{1-b} e^{\mp \pi i (1-b)}}
\\
\cdot\ & u^{1/2}
  \left[
      e^{\mp \frac{\pi i}{2} b} J_{-b}(\sqrt{\alpha} u)
    + \mu e^{\mp \pi i (1-b)} \alpha^{1-b}
      e^{\mp \frac{\pi i}{2} b} J_b   (\sqrt{\alpha} u)
  \right]
\\
\cdot\ & v^{1/2}
  \left[
      e^{\mp \frac{\pi i}{2} b} J_{-b}(\sqrt{\alpha} v)
    - e^{\mp \frac{\pi i}{2}b } J_b   (\sqrt{\alpha} v)
  \right].
\end{align*}
It is now easy to calculate the discontinuity 
$G_{\mu,\alpha + i0} -
G_{\mu,\alpha - i0}$ and to get for $\alpha=E \in \RR_+$
\begin{equation}
\ \quad  dE_\mu (u,v)
\ =\ 
\frac{\dd E}{2(1- 2 \mu E^{1-b} \cos b\pi +\mu^2 E^{2(1-b)})}
\,\varphi_{\mu,E}(u)\,\varphi_{\mu,E}(v)
\end{equation}
where $\varphi_{\mu,E}(u)$ given by \eqref{eq:eig-fct}
is the (generalized) eigenfunction 
of $N_\mu$ associated with the eigenvalue $E$. This corresponds
to the scalar spectral measure \eqref{eq:scal-spec-m}.

\section{Conservation of probability}
\label{apx:cons-proba}

We shall show here that the transition measures
$P_{\tilde\lambda}^t(r;\dd r')$ of the sticky process given by
\eqref{eq:trans-proba} are normalized, i.e.\,\,that $\int
P^t_{\tilde\lambda}(r,\dd r') = 1$.  This is clearly the case for
$t=0$ since $P_{\tilde\lambda}^0(r;\dd r') = \delta(r-r') \,\dd r'$.
But
$$
  \frac{\dd}{\dd t}
  \int e^{tM_{\tilde\lambda}}(r,r') \,\dd r'
=
  \int M^\dagger(r') \, e^{tM_{\tilde\lambda}}(r,r') \,\dd r'
=
  -F(r')
  \left[
    e^{tM_{\tilde\lambda}}(r,r')\,m(r')^{-1}
  \right]_{{r'=0}}
$$
where $M^\dagger = m(r) M \circ m(r)^{-1} = \partial_r \circ F \circ
m(r)^{-1}$ with $m(r) = r^{a_\xi-\xi}$ and $F = \tilde D_0 r^{a_\xi}
\partial_r$ is the formal adjoint of $M$ w.r.t. the $L^2(dr)$
scalar product.  The expression on the right hand side has the
interpretation of the mean flux of the probability current through
$r'=0$.  It should be balanced by the rate of change of probability to
stay at $r'=0$.  Using the relations
$$
  e^{tM_{\tilde\lambda}}(r,r') \, m(r')^{-1}
=
  e^{tM_{\tilde\lambda}}(r',r) \, m(r)^{-1}
$$
and the fact that $e^{tM_{\tilde\lambda}}(r,r')$ satisfies as a
function of $r$ the boundary condition \eqref{expr:boundary-cond}, we
infer that
\begin{align*}
  \frac{\dd}{\dd t}
  \int e^{tM_{\tilde\lambda}}(r,r') \,\dd r'
&=
  -\tilde\lambda \,
  M(r') \, e^{tM_{\tilde\lambda}}(r',r) \, m(r)^{-1}\Big|_{{r'=0}}
\\
&=
  -\tilde\lambda \, M(r')
  \Big(
    e^{tM_{\tilde\lambda}}(r,r') \, m(r')^{-1}
  \Big)\Big|_{{r'=0}}
\\
&=
  -\tilde\lambda \, m(r')^{-1}
  M^\dagger(r') \, e^{tM_{\tilde\lambda}}(r,r')\Big|_{{r'=0}}
\\
&=
  -\tilde\lambda \, \frac{\dd}{\dd t} \, m(r')^{-1} \,
  e^{tM_{\tilde\lambda}}(r,r')\Big|_{{r'=0}}
\,.
\end{align*}
It follows then from \eqref{eq:trans-proba} that the time derivative
of $\int P_{\tilde\lambda}^t(r;\dd r')$ vanishes so that the
normalization of $P_{\tilde\lambda}^t(r;\dd r')$ does not change in
time.

\section{Estimates of functions $f_1$ and $f_2$}
\label{apx:f1-f2}

We present bounds showing in which sense the approximations given for
functions $f_1$ and $f_2/f_1$ in \tabref{tab:f1-f2} are correct.
\vskip 0.15cm

The functions $f_{1,2}(r;l_\nu,l_\kappa)$ were defined by
\eqref{eq:f_1} and \eqref{eq:f_2} with $\kappa$ given by
\eqref{expr:kappa}. Let us examine their behavior for $\kappa = 0$
setting $f_{1,2}(r;l_\nu) \equiv f_{1,2}(r;l_\nu,0)$.  Let us first
show that $f_{1,2}(r;l_\nu)$ are positive for $r > 0$.  Note that
\eqref{def:d-reg} implies that
$$
  d_{ij}(\vec{r};l_\nu)
\,=\,
  D_0 \int \frac{2 \sin^2 \frac{\vec{k} . \vec{r}}{2}}
                {|\vec{k}|^{d + \xi}}
           \vec{\vec{P}}(\vec{k},\wp)
           f(l_\nu|\vec{k}|)
           \frac{\dd \vec{k}}
                {(2\pi)^d}
\,=\,\tilde D_0
  \int T_{ij}(\vec{k})\, \dd \vec{k}
$$
where $T(\vec{k})$ is a positive ($\vec{r}$-dependent) matrix.  On the
other hand, \eqref{eq:f_1} and \eqref{eq:f_2} may be written as
\begin{equation}
  f_1(r)
\,=\,
  \int T_{ij}(\vec{k})\,\frac{{r_i r_j}}{{r^2}}\, 
\dd \vec{k}\,, 
\qquad
  f_2(r)
\,=\,
  r^{-1}\int T_{ij}(\vec{k})\,(\delta_{ij}-\frac{{r_i r_j}}
{{r^2}})\,\dd\vec{k}
\nonumber
\end{equation}
for any vector $\vec{r}$ such that $|\vec{r}| = r$.  Positivity of $T$
implies
$$
  T_{ij}(\vec{k})\, r_i r_j
\geq
  0
\,,\qquad
  T_{ij}(\vec{k}) (\delta_{ij}-\frac{_{r_i r_j}}{^{r^2}})
\geq
  0\
$$
and for $r>0$ it is not hard to see that strict equalities hold except
for a set of $\vec{k}$ of measure zero.  Hence positivity of
$f_{1,2}(r;l_\nu)$.
\vskip 0.15cm

In order to study the behavior of $f_{1,2}(r;l_\nu)$ for small and
large $r$ it will be convenient to express them in terms of the function
\begin{equation}
g(r;l_\nu)\,=\,\int\frac{1-e^{i\vec{k}\cdot\vec{r}}}{|\vec{k}|^{d+\xi}}\,
f(l_\nu|\vec{k}|)\,\frac{\dd\vec{k}}{(2\pi)^d}\,.
\nonumber
\end{equation}
Note that $g$ is a smooth even function (i.e.\ all its derivatives of
odd order vanish at 0) of $r \geq 0$.  By scaling,
\begin{equation}
g(r;l_\nu)\,=\,l_\nu^\xi\,g(\frac{_r}{^{l_\nu}};1)\,.
\nonumber
\end{equation}
Around zero
\begin{equation}
\label{eq:ar0}
g(\rho;1)=C_0\rho^2\,+\,O(\rho^4)\qquad{\rm with}\qquad
C_0=\int\frac{f(|\vec{k}|)}{|\vec{k}|^{d+\xi-2}}\,
\frac{\dd\vec{k}}{(2\pi)^d}\,.
\end{equation}
Rewriting
$$
g(\rho;1)\,=\,\rho^\xi\int\frac{1-e^{i\vec{k}\cdot\vec{\rho}/\rho}}
{|\vec{k}|^{d+\xi}}\,f(|\vec{k}|/\rho)\,\frac{\dd\vec{k}}{(2\pi)^d}\,,
$$
we infer that around infinity
\begin{equation}
\label{eq:arinf}
g(\rho;1)\,=\,C_\infty\rho^\xi\,+\,O(\rho^{\xi-2})
\end{equation}
with 
\begin{equation}
C_\infty\,=\,\int\frac{1-e^{i\vec{k}\cdot\vec{\rho}/\rho}}
{|\vec{k}|^{d+\xi}}\,\frac{\dd\vec{k}}{(2\pi)^d}\,=\,
\frac{\Gamma(\frac{2-\xi}{2})}{2^{d+\xi-2}\pi^{\frac{d}{2}}\xi\,
\Gamma(\frac{d+\xi}{2})}\,.
\nonumber
\end{equation}
The spatial covariance $\vec{\vec{d}}(\vec{r};l_\nu)$ given by
\eqref{def:d-reg} may be expressed in terms of the function $g$:
\begin{multline}
\label{eq:dg}
\ \hspace{2cm}d_{ij}(\vec{r};l_\nu)
\,=\,D_0\Big[\Big(\frac{1-\wp}{d-1}\delta_{ij}
+\frac{\wp d-1}{d-1}\frac{r_ir_j}{r^2}\Big)\,g(r;l_\nu)
\\
+\,\frac{\wp d-1}{d-1}\left(\delta_{ij}-d\frac{r_ir_j}{r^2}\right)r^{-d}
\int\limits_0^r(r')^{\,d-1}g(r';l_\nu)\,\dd r'\Big].\hspace{1.25cm}
\end{multline}
Indeed, by rotational covariance, $d_{ij}$ has to be a combination of
$\,\delta_{ij}$ and $\frac{r_ir_j}{r^2}$. To show that the coefficient
functions are correctly represented above, it is enough to note that
$d_{ii}$ and $\partial_id_{ij}$ are, in virtue of \eqref{def:d-reg},
expressible in terms of $g$ and of its $r$-derivative $g'$ as
$d_{ii}=D_0g$ and $\partial_id_{ij} =D_0\wp\frac{r_j}{r}g'$ and that
the same relations may be recovered from \eqref{eq:dg}.  Now using
definitions \eqref{eq:f_1} and \eqref{eq:f_2} we obtain:
\begin{gather}
\label{eq:f1g}
  \frac{_{\tilde{D}_0}}{^{D_0}}\,f_1(r;l_\nu)
\,=\,
    \frac{1-\wp}{d-1}\,g(r;l_\nu) \,+\,
    \frac{\wp d-1}{d-1}
    \left(
      g(r;l_\nu)-
      \frac{d-1}{r^d} \int_0^r{r'}^{\,d-1} \,g(r';l_\nu) \,\dd r'
    \right),
\\
\label{eq:f2g}
    \frac{_{\tilde{D}_0}}{^{D_0}}\,f_2(r;l_\nu)
\,=\,
    \frac{1-\wp}{r}\,g(r;l_\nu) \,+\,
    \frac{\wp d-1}{r^d} \int_0^r{r'}^{\,d-1} \,g(r';l_\nu)\,\dd r'
\,.\hspace{2.9cm}
\end{gather} 
Functions $f_{1,2}(r;l_\nu)$ inherit from $g(r;l_\nu)$ the scaling
property \eqref{eq:scal12}.  Relations \eqref{eq:f1g}, \eqref{eq:f2g}
and expansions \eqref{eq:ar0} and \eqref{eq:arinf} permit now to write
the decompositions \eqref{ga:gad1} to \eqref{ga:gad4} with $\tilde
C_0=\frac{1+2\wp}{d+2}C_0D_0/\tilde D_0$ and similar coefficient in
\eqref{ga:gad3} and \eqref{ga:gad4} fixed to $1$ by setting ${\tilde
D}_0=\frac{1+\wp\xi}{d+\xi}C_\infty D_0$.  Moreover the positivity of
functions $f_{1,2}(r;1)$ and their smoothness imply that there exists
some constant $B > 0$ such that for all $0 \leq \rho \leq 1$
\begin{gather}
\label{ga:bd1}
  \frac{1}{B}
<
  f_i^\clubsuit(\rho)
<
  B
\qquad\,\, i=1,2\,,
\\
\label{ga:bd2}
  -B
<
  f_i^\spadesuit(\rho)
<
  B
\qquad i=1,2
\end{gather}
and for all $1 \leq \rho < \infty$,
\begin{gather}
  \frac{1}{B}
\label{ga:bd3}
<
  f_i^\diamondsuit(\rho)
<
  B
\qquad\,\, i=1,2\,,
\\
\label{ga:bd4}
  -B
<
  f_i^\heartsuit(\rho)
<
  B
\qquad i=1,2\,.
\end{gather}
\vskip 0.15cm

Let us come back to functions $f_{1,2}(r;l_\nu,l_\kappa)$ with 
$0<l_\kappa,l_\nu<1$. To obtain bounds for $f_1$, we write for $l_\nu < r$
$$
  f_1(r; l_\nu,l_\kappa)
=
  r^\xi f_1^\diamondsuit(\frac{r}{l_\nu})
  + 2l_\nu^{\xi-2} l_\kappa^2
$$
and infer that
\begin{equation}
\label{bd:f_1;1}
  r^\xi \frac{1}{B}
<
  f_1(r; l_\nu,l_\kappa)
<
  r^\xi (B+2)
\,.
\end{equation}
Similarly, the decomposition
$$
  f_1(r; l_\nu,l_\kappa)
=
  \tilde C_0\, l_\nu^{\xi-2} r^2 f_1^\clubsuit(\frac{r}{l_\nu})
  + 2l_\nu^{\xi-2} l_\kappa^2
$$
gives for $l_\kappa < r < l_\nu$ the bounds
\begin{equation}
  l_\nu^{\xi-2} r^2 \frac{\tilde C_0}{B}
<
  f_1(r; l_\nu,l_\kappa)
<
  l_\nu^{\xi-2} r^2 (\tilde C_0 B+2)
\nonumber
\end{equation}
and for $r < l_\kappa$,
\begin{equation}
\label{bd:f_1;3}
  2l_\nu^{\xi-2} l_\kappa^2
<
  f_1(r; l_\nu,l_\kappa)
<
  l_\nu^{\xi-2} l_\kappa^2 (\tilde C_0 B+2)
\,.
\end{equation}
This shows that $f_1(r;l_\nu,l_\kappa)$ is bounded above and below by
its approximate version of \tabref{tab:f1-f2} multiplied or divided by
a constant.
\vskip 0.15cm

Function $f_2$ may be estimated similarly resulting in $O(1)$ bounds
for $rf_2/f_1$.  We need however (in particular in \appref{apx:m(0)})
the following more precise estimates for that ratio:
\begin{alignat}{3}
\label{bd:f_2/f_1;1}
  &\text{for $l_\nu < r$~:}
\qquad
  &&\left| \frac{f_2(r;l_\nu,l_\kappa)}{f_1(r;l_\nu,l_\kappa)}
  - \frac{a_\xi}{r}\right|
&&\quad<\ \ 
  C \frac{l_\nu^\xi}{r^{1+\xi}}
\\
\label{bd:f_2/f_1;2}
  &\text{for $l_\kappa < r < l_\nu$~:}
\qquad
  &&\left| \frac{f_2(r;l_\nu,l_\kappa)}{f_1(r;l_\nu,l_\kappa)}
  - \frac{a_2}{r}\right|
&&\quad<\ \ 
  C \left[\frac{r}{l_\nu^2} + \frac{l_\kappa^2}{r^3}\right]
\\
\label{bd:f_2/f_1;3}
  &\text{for $r < l_\kappa$~:}
\qquad\quad
  &&\left| \frac{f_2(r;l_\nu,l_\kappa)}{f_1(r;l_\nu,l_\kappa)}
  - \frac{d-1}{r}\right|
&&\quad<\ \ 
  C \frac{r}{l_\kappa^2 }
\end{alignat}
for some constant $C$ which may be chosen independent of
$r,l_\nu,l_\kappa$.  To establish \eqref{bd:f_2/f_1;1} we use the
decompositions \eqref{ga:gad3} and \eqref{ga:gad4} to write for $l_\nu
< r$
\begin{align*}
 &\left|
    \frac{f_2(r;l_\nu,l_\kappa)}{f_1(r;l_\nu,l_\kappa)}
    - \frac{a_\xi}{r}
  \right|
\ =\ 
\left|
    \frac{
      l_\nu^{\xi-1} f_2(\frac{r}{l_\nu};1)
      + 2\frac{d-1}{r} l_\nu^{\xi-2} l_\kappa^2
      - \frac{a_\xi}{r}
        \left[l_\nu^\xi f_1(\frac{r}{l_\nu};1)
              + 2l_\nu^{\xi-2} l_\kappa^2
        \right]
    }{
      l_\nu^\xi f_1(\frac{r}{l_\nu};1)
      + 2l_\nu^{\xi-2} l_\kappa^2
    }
  \right|
\\
&=\ 
  \left|
    \frac{
      l_\nu^2 r^{\xi-3}
      \left[
        f_2^\heartsuit(\frac{r}{l_\nu})
        - a_\xi f_1^\heartsuit(\frac{r}{l_\nu})
      \right]
      + 2\frac{d-1-a_\xi}{r} l_\nu^{\xi-2} l_\kappa^2
    }{
      r^\xi
      f_1^\diamondsuit(\frac{r}{l_\nu})
      + 2l_\nu^{\xi-2} l_\kappa^2
    }
  \right|
\ <\ 
    \frac{B [1+|a_\xi|] + 2|d-1-a_\xi|}
       {\frac{1}{B}}
    \,\frac{l_\nu^\xi}{r^{1+\xi}}
\end{align*}
where the inequality follows from 
the bounds \eqref{ga:bd3} and \eqref{ga:bd4}.
In the same manner for $l_\kappa < r < l_\nu$,
using the decompositions \eqref{ga:gad1},
\eqref{ga:gad2} and the bounds \eqref{ga:bd1} and \eqref{ga:bd2},
we obtain
\begin{align*}
  \left|
    \frac{f_2(r;l_\nu,l_\kappa)}{f_1(r;l_\nu,l_\kappa)}
    - \frac{a_2}{r}
  \right|
&=
  \left|
    \frac{
      l_\nu^{\xi-4} r^3
      \left[
        f_2^\spadesuit(\frac{r}{l_\nu})
        - a_2 f_1^\spadesuit(\frac{r}{l_\nu})
      \right]
      + 2\frac{d-1-a_2}{r} l_\nu^{\xi-2} l_\kappa^2
    }{
      \tilde C_0 l_\nu^{\xi-2} r^2
      f_1^\clubsuit(\frac{r}{l_\nu})
      + 2l_\nu^{\xi-2} l_\kappa^2
    }
  \right|
\\
&<
  \left[\frac{r}{l_\nu^2} + \frac{l_\kappa^2}{r^3}\right]
  \frac{B [1+|a_2|] + 2|d-1-a_2|}
       {\frac{\tilde C_0}{B}}\,.
\end{align*}
Finally for $r < l_\kappa$,
\begin{align*}
  \left|
    \frac{f_2(r;l_\nu,l_\kappa)}{f_1(r;l_\nu,l_\kappa)}
    - \frac{d-1}{r}
  \right|
&\ =\ 
  \left|
    \frac{
      \tilde C_0(a_2-d+1)l_\nu^{\xi-2}\,r+ 
      l_\nu^{\xi-4} r^3
      \left[
        f_2^\spadesuit(\frac{r}{l_\nu})
        - (d-1) f_1^\spadesuit(\frac{r}{l_\nu})
      \right]
    }{
      \tilde C_0 l_\nu^{\xi-2} r^2
      f_1^\clubsuit(\frac{r}{l_\nu})
      + 2l_\nu^{\xi-2} l_\kappa^2
    }
  \right|
\\
&\ <\ 
  \frac{r}{l_\kappa^2}
  \,\frac{\tilde C_0|d-1-a_2|+B [1+|d-1|]}{2}\,.
\end{align*}
\vskip 0.15cm

Let us end this Appendix by listing the scaling forms of the tensor
$\,\vec{\vec{d}}\,$ and of functions $\,f_{1,2}$.  Since
$\,  g(r;0)=  C_\infty\,r^\xi$, \,we infer from \eqref{eq:dg} that
$$
  d_{ij}(\vec{r},0)
\,=\,
  \frac{\tilde D_0}{d-1}
  \Big[
    \Big(
      \frac{d+\xi}{1+\xi\wp}-1
    \Big)
    \delta_{ij} +
    \frac{\xi(\wp d-1)}{1+\xi\wp} \,
    \frac{r_ir_j}{r^2}
  \Big]
  \,r^\xi
$$
and that $f_1(r;0) = r^\xi$ and $f_2(r;0) = a_\xi\,r^{\xi-1}$, in
agreement with the scaling form \eqref{eq:M} of the generator $M$.

\section{Exact speed measure at zero}
\label{apx:m(0)}

Here we prove the convergence \eqref{eq:m(0)}, establishing the exact
value of the mass at zero of the limiting speed measure. Using 
\eqref{eq:speed-meas}), we have
$$
  \int_0^{l_\nu}
    m(r';l_\nu,l_\kappa)\, \chi(r')
  \,\dd r'
\,=\,
  \int_0^{l_\nu}
    \frac{\chi(r')}{f_1(r';l_\nu,l_\kappa)}
    \,\exp\left(
      -\int_{r'}^{r_0}
         \frac{f_2(r'';l_\nu,l_\kappa)}
              {f_1(r'';l_\nu,l_\kappa)}
         \,\dd r''
    \right)
  \,\dd r'\,.
$$
The integral in the exponential may be split into four terms:
\begin{multline*}
  \int_{r'}^{r_0}
    \frac{f_2(r'';l_\nu,l_\kappa)}
         {f_1(r'';l_\nu,l_\kappa)}
    \,\dd r''
\ =\ 
  \int_{r'}^{l_\kappa}
    \frac{f_2(r'';l_\nu,l_\kappa)}
         {f_1(r'';l_\nu,l_\kappa)}
    \,\dd r''
\\
 +\, \int_{l_\kappa}^{\sqrt{l_\kappa l_\nu}}
    \frac{f_2(r'';l_\nu,l_\kappa)}
         {f_1(r'';l_\nu,l_\kappa)}
    \,\dd r''
  \,+\,
  \int_{\sqrt{l_\kappa l_\nu}}^{l_\nu}
    \frac{f_2(r'';l_\nu,l_\kappa)}
         {f_1(r'';l_\nu,l_\kappa)}
    \,\dd r''
  \,+\,
  \int_{l_\nu}^{r_0}
    \frac{f_2(r'';l_\nu,l_\kappa)}
         {f_1(r'';l_\nu,l_\kappa)}
    \,\dd r''\,.\quad
\end{multline*}
We begin by calculating the last three terms as they do not depend on
$r'$.  The last one gives
\begin{multline*}
\hspace{-0.2cm}
 \int_{l_\nu}^{r_0}
    \frac{f_2(r'';l_\nu,l_\kappa)}
         {f_1(r'';l_\nu,l_\kappa)}
    \,\dd r''\,
  =\, a_\xi \ln \frac{r_0}{l_\nu}
  \,+
  \int_1^{\frac{r_0}{l_\nu}}
    \frac{
      \rho^{\xi-3}\hspace{-0.07cm}
      \left[
        f_2^\heartsuit(\rho)
        - a_\xi f_1^\heartsuit(\rho)
      \right]\hspace{-0.07cm}
      +
      2\frac{d-1-a_\xi}{\rho}
      \left(\frac{l_\kappa}{l_\nu}\right)^2
    }{
      \rho^\xi
      f_1^\diamondsuit(\rho)
      + 2\left(\frac{l_\kappa}{l_\nu}\right)^2
    }
  \,\dd \rho
\end{multline*}
and we know from \eqref{bd:f_2/f_1;1} that the integrand on the right
is dominated by $C \rho^{-(1+\xi)}$ (as soon as $l_\nu < 1$), so we
may use the Dominated Convergence Theorem to conclude that
$$
  \lim_{l_\nu,\frac{l_\kappa}{l_\nu} \rightarrow 0} \left(
    \int_{l_\nu}^{r_0}
      \frac{f_2(r'';l_\nu,l_\kappa)}
           {f_1(r'';l_\nu,l_\kappa)}
      \,\dd r''
    -
    a_\xi \ln \frac{r_0}{l_\nu}
  \right)
\ =\ 
  \int_1^\infty
    \frac{
      f_2^\heartsuit(\rho)
      - a_\xi f_1^\heartsuit(\rho)
    }{
      \rho^3
      f_1^\diamondsuit(\rho)
    }
  \,\dd \rho\,.
$$
As for the other two,
\begin{multline*}
  \int_{l_\kappa}^{\sqrt{l_\kappa l_\nu}}
    \frac{f_2(r'';l_\nu,l_\kappa)}
         {f_1(r'';l_\nu,l_\kappa)}
    \,\dd r''
\\
  =\  a_2\, \ln \sqrt{\frac{l_\nu}{l_\kappa}}
  +
  \int_{1}^{\sqrt{l_\nu/l_\kappa}}
    \frac{
      \left(\frac{l_\kappa}{l_\nu}\right)^2 \rho^3
      \left[
        f_2^\spadesuit(\frac{l_\kappa}{l_\nu} \rho)
        - a_2 f_1^\spadesuit(\frac{l_\kappa}{l_\nu} \rho)
      \right]
      + 2\frac{d-1-a_2}{\rho}
    }{
      \tilde C_0 \rho^2
      f_1^\clubsuit(\frac{l_\kappa}{l_\nu} \rho)
      + 2
    }
  \,\dd \rho\quad
\end{multline*}
with the integrand on the right dominated by 
$C [(\frac{l_\kappa}{l_\nu})^2 \rho + \rho^{-3}] < 2C
\rho^{-3}$, see  \eqref{bd:f_2/f_1;2}, so we may
conclude that
\begin{multline*}
  \lim_{l_\nu,\frac{l_\kappa}{l_\nu} \rightarrow 0} \left(
    \int_{l_\kappa}^{\sqrt{l_\kappa l_\nu}}
      \frac{f_2(r'';l_\nu,l_\kappa)}
           {f_1(r'';l_\nu,l_\kappa)}
      \,\dd r''
    -
    a_2 \ln \sqrt{\frac{l_\nu}{l_\kappa}}
  \right)
\\
  =\ \int_{1}^{\infty}
    \frac{2(d-1-a_2)}
         {\rho\,(\tilde C_0 \rho^2 + 2)}
  \,\dd \rho
\ =\ 
  \frac{d-1-a_2}{2}
  \ln \frac{\tilde C_0 + 2}{\tilde C_0}\,.\quad
\end{multline*}
Similarly,
\begin{multline*}
  \ \ \int_{\sqrt{l_\kappa l_\nu}}^{l_\nu}
    \frac{f_2(r'';l_\nu,l_\kappa)}
         {f_1(r'';l_\nu,l_\kappa)}
    \,\dd r''
\\
 =\  a_2 \ln \sqrt{\frac{l_\nu}{l_\kappa}}
  \,+\,
  \int_{\sqrt{l_\kappa/l_\nu}}^{1}
    \frac{
      \rho^3
      \left[
        f_2^\spadesuit(\rho)
        - a_2 f_1^\spadesuit(\rho)
      \right]
      + 2\frac{d-1-a_2}{\rho} \left(\frac{l_\kappa}{l_\nu}\right)^2
    }{
      \tilde C_0 \rho^2
      f_1^\clubsuit(\rho)
      + 2\left(\frac{l_\kappa}{l_\nu}\right)^2 
    }
  \,\dd \rho\quad
\end{multline*}
where the integrand on the right is bounded by 
$C [\rho + (\frac{l_\kappa}{l_\nu})^2 \rho^{-3}] < 2C
\rho$, see \eqref{bd:f_2/f_1;2}. Hence
$$
  \lim_{l_\nu,\frac{l_\kappa}{l_\nu} \rightarrow 0} \left(
    \int_{\sqrt{l_\kappa l_\nu}}^{l_\nu}
      \frac{f_2(r'';l_\nu,l_\kappa)}
           {f_1(r'';l_\nu,l_\kappa)}
      \,\dd r''
    -
    a_2 \ln \sqrt{\frac{l_\nu}{l_\kappa}}
  \right)
=
  \int_{0}^{1}
    \frac{
      \rho
      \left[
        f_2^\spadesuit(\rho)
        - a_2 f_1^\spadesuit(\rho)
      \right]
    }{
      \tilde C_0
      f_1^\clubsuit(\rho)
    }
  \,\dd \rho\,.\quad
$$
What remains to be evaluated is
\begin{multline*}
  \int_0^{l_\nu}
    \frac{\chi(r')}{f_1(r';l_\nu,l_\kappa)}
    \exp\left(
      -\int_{r'}^{l_\kappa}
         \frac{f_2(r'';l_\nu,l_\kappa)}
              {f_1(r'';l_\nu,l_\kappa)}
         \,\dd r''
    \right)
  \,\dd r'
\\
  =\ \frac{1}{l_\nu^{\xi-2} l_\kappa}
  \int_0^{l_\nu/l_\kappa}
    \frac{\chi(l_\kappa \rho)}
         {\tilde C_0 \rho^2 f_1^\clubsuit(\frac{l_\kappa}{l_\nu} \rho)
          + 2
         }
    \,\exp\left[
      - \int_\rho^1
          \frac{
            \tilde C_0 a_2 \rho' f_2^\clubsuit(\frac{l_\kappa}{l_\nu} \rho')
            +2\frac{d-1}{\rho'}
          }{
            \tilde C_0 {\rho'}^2 f_1^\clubsuit(\frac{l_\kappa}{l_\nu} \rho')
            + 2
          }
        \,\dd \rho'
    \right]
  \,\dd \rho\,.\quad
\end{multline*}
Let us show that the Dominated Convergence Theorem applies once more.  
The following estimates are sufficient.  For $0 < \rho < 1$,
$$
  \left|
    \frac{\chi(l_\kappa \rho)}
         {\tilde C_0 \rho^2 f_1^\clubsuit(\frac{l_\kappa}{l_\nu} \rho)
          + 1
         }
  \right|
<
  \chi(0)\,,\qquad
  \left|
    \frac{
      \tilde C_0 a_2 \rho f_2^\clubsuit(\frac{l_\kappa}{l_\nu} \rho)
      +\frac{d-1}{\rho}
    }{
      \tilde C_0 {\rho}^2 f_1^\clubsuit(\frac{l_\kappa}{l_\nu} \rho)
      + 1
    }
    - \frac{d-1}{\rho}
  \right|
<\,C\rho\,.
$$
The first inequality is a consequence of $\chi(r) < \chi(0)$ 
for any $r > 0$, and of the positivity of $f_1^\clubsuit$.  
The second one is obtained
by rewriting \eqref{bd:f_2/f_1;3} using \eqref{ga:gad1} and
\eqref{ga:gad2}.
\,For $1 < \rho < l_\nu/l_\kappa$, we use
$$
  \left|
    \frac{\chi(l_\kappa \rho)}
         {\tilde C_0 \rho^2 f_1^\clubsuit(\frac{l_\kappa}{l_\nu} \rho)
          + 1
         }
  \right|
<
  \frac{\chi(0)}
       {\frac{\tilde C_0}{B} \rho^2}\,,\qquad
  \left|
    \frac{
      \tilde C_0 a_2 \rho f_2^\clubsuit(\frac{l_\kappa}{l_\nu} \rho)
      +\frac{d-1}{\rho}
    }{
      \tilde C_0 {\rho}^2 f_1^\clubsuit(\frac{l_\kappa}{l_\nu} \rho)
      + 1
    }
    - \frac{a_2}{\rho}
  \right|
<
  C \left[
      \left(\frac{l_\kappa}{l_\nu}\right)^2 \rho
      + \frac{1}{\rho^3}
    \right].
$$
Here, the first inequality follows from \eqref{ga:bd1} and the second one is
obtained by rewriting \eqref{bd:f_2/f_1;2} with the use of \eqref{ga:gad1} 
and \eqref{ga:gad2}. \,We then have
\begin{multline*}
\hspace{1cm}
  \lim_{l_\nu,\frac{l_\kappa}{l_\nu} \rightarrow 0}
    l_\nu^{\xi-2} l_\kappa
    \int_0^{l_\nu}
      \frac{\chi(r')}{f_1(r';l_\nu,l_\kappa)}
      \exp\left(
        -\int_{r'}^{l_\kappa}
           \frac{f_2(r'';l_\nu,l_\kappa)}
                {f_1(r'';l_\nu,l_\kappa)}
           \,\dd r''
      \right)
    \,\dd r'
\\
\hspace{1cm}
   =\, \int_0^\infty
    \frac{\chi(0)}
         {\tilde C_0 \rho^2
          + 2
         }
    \exp\left[
      - \int_\rho^1
          \frac{
            \tilde C_0 a_2 \rho'
            +2\frac{d-1}{\rho'}
          }{
            \tilde C_0 {\rho'}^2
            + 1
          }
        \,\dd \rho'
    \right]
  \,\dd \rho
\\
\hspace{1cm}
  =\, 2^{\frac{a_2-3}{2}}\,
  (\tilde C_0+2)^\frac{d-1-a_2}{2}\,{\tilde C}_0^{-\frac{d}{2}}
  \,B(\frac{_d}{^2},\frac{_{1-a_2}}{^{2}})
  \,\chi(0)\hspace{1.9cm}
\end{multline*}
where $B(x,y)$ is the Euler Beta function.  Gathering all terms, in
particular the powers of $\,l_\nu$ and $l_\kappa$ that combine to
$l_\nu^{(a_\xi+1-\xi)+(1-a_2)} l_\kappa^{a_2-1}$, we obtain the result
\eqref{eq:m(0)} with $Y$ given by \eqref{eq:Y}.
\vskip -0.1cm
\


\begin{thebibliography}{bib}

\bibitem{SM}
Bernard, D., Gaw\c{e}dzki, K., Kupiainen, A.:
``Slow modes in passive advection'', J. Stat. Phys. {\bf 90}
(1998), 519-569
\vskip 0.15cm

\bibitem{HBM}
Borodin, A., Salminen, P.: ``Handbook of Brownian Motion: Facts and
Formulae'', Birkh\"{a}user, Boston 1996
\vskip 0.15cm

\bibitem{Breiman}
Breiman, L.: ``Probability'', Addison-Wesley, Reading MA 1968
\vskip 0.15cm

\bibitem{CHGKV}
Chaves, M., Horvai, P., Gaw\c{e}dzki, K., Kupiainen, A., Vergassola, M.:
``Lagrangian dispersion in Gaussian self-similar ensembles'',
arXiv:nlin.CD/0303031, to appear in J. Stat. Phys. 
\vskip 0.15cm

\bibitem{EVDE1}
E, W., Vanden-Eijnden, E.: ``Generalized flows, intrinsic
stochasticity, and turbulent transport'', Proc. Natl. Acad. Sci.
USA, {\bf 97} (2000), 8200-8205
\vskip 0.15cm

\bibitem{EVDE2}
E, W., Vanden-Eijnden, E.: ``Turbulent Prandtl number effect on
passive scalar advection'', Physica D {\bf 152-153} (2001),
636-645
\vskip 0.15cm

\bibitem{FGV}
Falkovich, G., Gaw\c{e}dzki, K., Vergassola, M.: ``Particles and
fields in fluid turbulence'', Rev. Mod. Phys. {\bf 73} (2001),
913-975
\vskip 0.15cm

\bibitem{Feller}
Feller, W.: ``The parabolic differential equations and 
the associated semi-groups of transformations'', 
Ann. Math. {\bf 55}, 468-519 (1952)
\vskip 0.15cm

\bibitem{FW}
Freidlin, M. I., Wentzell, A. D.: ``Necessary and Sufficient Conditions
for Weak Convergence of One-Dimensional Markov Processes'', 
in: ``The Dynkin Festschrift, Markov processes and their Applications''
(ed.\ Freidlin,~M.I.), pp.~95-109, Birk\-h\"{a}u\-ser, Boston, 1994
\vskip 0.15cm

\bibitem{G}
Gaw\c{e}dzki, K., unpublished
\vskip 0.15cm

\bibitem{GV}
Gaw\c{e}dzki, K., Vergassola, M.: ``Phase transition in the passive 
scalar advection'', Physica D {\bf 138} (2000), 63-90
\vskip 0.15cm

\bibitem{GR}
Gradsteyn, I.S. Ryzhik, I.M.: ``Table of Integrals: Series, and Products''
Academic Press, New York 1980
\vskip 0.15cm

\bibitem{IMcK}
It\^{o}, K., McKean, H. P.: ``Diffusion Processes and Their Sample Paths'',
Springer, Berlin 1965
\vskip 0.15cm

\bibitem{Ka68}
Kazantzev, A. P.: ``Enhancement of a magnetic field by a conducting fluid''.
Sov. Phys. JETP {\bf 26} (1968), 1031-1034
\vskip 0.15cm

\bibitem{Kr68}
Kraichnan, R. H.: ``Small-scale structure of a scalar field convected
by turbulence'', Phys. Fluids {\bf 11} (1968), 945-963
\vskip 0.15cm

\bibitem{LeJR1}
Le Jan, Y., Raimond, O.:  ``Integration of Brownian vector
fields'', Ann. Probab. {\bf 30} (2002), 826-873
\vskip 0.15cm

\bibitem{LeJR2}
Le Jan, Y., Raimond, O.:  ``Flows, coalescence and noise'',
arXiv:math.PR/0203221, to appear in Ann. Probab.
\vskip 0.15cm

\bibitem{LeJR3}
Le Jan, Y., Raimond, O.:  ``Sticky flows on the circle'',
arXiv:math.PR/0211387 
\vskip 0.15cm

\bibitem{Onsager}
Onsager, L.: ``Statistical hydrodynamics''. Nuovo Cim. Suppl. 
{\bf 6} (1949), 279-287 
\vskip 0.15cm

\bibitem{RW}
Rogers, L. C. G., Williams, D.:
``Diffusions, Markov Processes and Martingales''.
Cambridge University Press, Cambridge 2000
\vskip 0.15cm

\bibitem{RY}
Revuz, D., Yor, M.: ``Continuous Martingales and Brownian Motion'', 
Springer, Berlin 1991
\vskip 0.15cm

\end{thebibliography}
\end{document}